\documentclass[twocolumn,showpacs,pre,citeautoscript,amsmath,amssymb,floatfix,eqsecnum]{revtex4-1}
\usepackage{amsmath,amssymb,color}
\usepackage{bm}
\usepackage{graphicx}
\usepackage{psfrag}
\usepackage{mhchem}
\usepackage{comment}
\newcommand{\bd}{\bm}
\newcommand{\rt}{\tilde{\rho}}
\newcommand{\mb}{\bar{m}}
\newcommand{\tTheta}{\tilde{\vartheta}}

\begin{document}

\title{Dual lattice functional renormalization group for
the Berezinskii-Kosterlitz-Thouless
transition: irrelevance of amplitude and out-of-plane fluctuations}

\author{Jan Krieg and Peter Kopietz}
  
\affiliation{Institut f\"{u}r Theoretische Physik, Universit\"{a}t
  Frankfurt,  Max-von-Laue Strasse 1, 60438 Frankfurt, Germany}

%\date{\today}
\date{October 6, 2017}

 \begin{abstract}

We develop a new functional renormalization group (FRG) approach for the two-dimensional XY-model by combining the lattice FRG proposed by Machado and Dupuis [Phys. Rev. E {\bf{82}}, 041128 (2010)] with a duality transformation which explicitly introduces vortices via an integer-valued field. We show that the hierarchy of FRG flow equations for the infinite set of  relevant and marginal couplings of the model can be reduced to the well-known Kosterlitz-Thouless renormalization group equations for the renormalized temperature and the vortex fugacity. Within our approach it is straightforward to include weak amplitude as well as out-of-plane fluctuations of the spins, which lead to additional interactions between the vortices that do not spoil the Berezinskii-Kosterlitz-Thouless transition. This demonstrates that previous failures to obtain a line of true fixed points within the FRG are a mathematical artifact of insufficient truncation schemes.

\end{abstract}

%\pacs{XXXX}

\maketitle

\section{Introduction}
\label{sec:intro}

The discovery of the Berezinskii-Kosterlitz-Thouless (BKT) phase transition by Berezinskii \cite{Berezinskii71,Berezinskii72} and by Kosterlitz and Thouless \cite{Kosterlitz72,Kosterlitz73} was important for several reasons. First of all, it demonstrated that superconductivity and superfluidity are possible at finite temperature $T$ even in two dimensions. Although true long-range order in two-dimensional systems with continuous symmetry is forbidden by the Mermin-Wagner theorem \cite{Mermin66}, it does not exclude the possibility of quasi-long-range order, which manifests itself in algebraically decaying correlations. Another remarkable property of the BKT transition is that it is a continuous phase transition which does not break any symmetry, which was unexpected at the time of its discovery. In fact, the insight that the BKT transition cannot be characterized by a broken symmetry but is driven by topological defects had a huge impact on future research in condensed matter physics, where nowadays topological concepts play a central role. In the XY-model, these topological defects correspond to vortex and anti-vortex configurations of the spins, which cannot be continuously deformed into the ferromagnetic ground state, making them local minima of the energy. While at temperatures $T$ below the critical temperature $T_c$ only bound pairs of vortices and anti-vortices are energetically allowed, this changes for $T > T_c$ where free vortices proliferate due to the accompanying entropy gain.

The discovery of the BKT transition was the beginning of a line of research where topological concepts are central to understand the behavior of condensed matter systems. Important milestones in this field are the explanation by Thouless {\it{et al.}} \cite{Thouless82} of the integer quantum Hall effect in terms of topology, the topological characterization of spin chains by Haldane \cite{Haldane83A,Haldane83B}, and Laughlin's \cite{Laughlin83} theoretical explanation of the fractional quantum Hall effect as a quantum liquid, which led to the discovery of topological order \cite{Wen90}. This novel kind of order, which is also found in quantum spin liquids \cite{Balents10} and in superconductors \cite{Hansson04}, is different from the conventional one described within Landau theory in that it is not related to any symmetries or to their breaking. Topological order refers to a topologically non-trivial ground state with long-range entanglement and is thus intrinsically a quantum effect. Its non-trivial topology makes the ground state robust against arbitrary local perturbations. This is in contrast to the related concept of symmetry protected topological order, which features short-range entanglement and is robust only against perturbations which preserve the underlying symmetry. The earliest example of symmetry protected topological order is given by the Haldane phase of a spin-1 chain, but it also appears, e.g., in topological insulators \cite{Hasan10} and in Weyl semimetals \cite{Xu15}.

In their original work on the BKT transition, Kosterlitz and Thouless \cite{Kosterlitz73} used a real-space renormalization group (RG) approach to calculate the critical properties of the two-dimensional XY-model \cite{Kosterlitz74,Young78}. The RG procedure invented by Kosterlitz and Thouless \cite{Kosterlitz73} is rather unconventional and supports the point of view that nothing is automatic about the RG, which often requires non-trivial reformulations adopted to the specific physical problem \cite{Herbut07}. The theoretical predictions of Kosterlitz and Thouless \cite{Kosterlitz73,Kosterlitz74} are in good agreement with experimental observations in several systems which are believed to exhibit a BKT transition, such as liquid $\ce{^{4}He}$-films \cite{Bishop78,Agnolet89}, arrays of Josephson junctions \cite{Resnick81}, or ultracold gases \cite{Hadzibabic06,Schweikhard07,Fletcher15}. However, the two-dimensional XY-model is not a fully accurate description of the experimental systems. In particular, the XY-model neither contains fluctuations of the length of the spins (amplitude fluctuations) nor does it allow for out-of-plane motion of the spins, which in real physical systems cannot be completely eliminated. Generally it is believed that amplitude fluctuations are innocuous in that they can be absorbed into a finite renormalization of the exchange coupling in an effective XY-model \cite{Nelson02}. This has recently been demonstrated explicitly by Erez and Meir \cite{Erez13} for the attractive Hubbard model using classical Monte Carlo simulations.

In recent years, the development of functional renormalization group (FRG) methods \cite{Wetterich93,Berges02,Kopietz10} has somewhat unified different formulations of the RG by providing a mathematically elegant and formally exact formulation of the Wilsonian idea of mode elimination and rescaling. However, the RG flow equations derived by Kosterlitz and Thouless \cite{Kosterlitz73} have not been recovered within the framework of the FRG. In fact, in a recent FRG study of the classical $O (2)$-model, Jakubczyk and Metzner \cite{Jakubczyk17} discussed the possibility that the effect of amplitude fluctuations on the BKT transition might be stronger than previously assumed and might even destroy the BKT transition. This point of view is supported by the fact that previous FRG calculations for the classical $O (2)$-model have so far not reproduced the line of fixed points describing the BKT phase in a satisfactory way. While signatures of the BKT transition have been seen, the finite mass of the amplitude fluctuations results in a line of quasi-fixed points so that the BKT transition is replaced by a smooth crossover \cite{Graeter95,Gersdorff01,Jakubczyk17}. Although it is possible to fine-tune the cutoff procedure for each temperature separately such that one encounters a true fixed point \cite{Jakubczyk14}, this fine-tuning can only be justified if one assumes that the quasi-fixed points are a mathematical artifact of the derivative expansion. Indeed, the lack of true fixed points without fine-tuning motivated Jakubczyk and Metzner \cite{Jakubczyk17} to consider the possibility that the line of quasi-fixed points correctly describes the physical system.
This is not in contradiction to experiments and to numerical simulations as the quasi-fixed points found in FRG calculations correspond to a finite, yet very large correlation length which might be difficult to distinguish at finite system sizes \cite{Jakubczyk17}.
A novel procedure to study the classical $O (2)$-model within the FRG was recently developed by Defenu {\it{et al.}} \cite{Defenu17}, where they decoupled phase and amplitude fluctuations by hand. This allowed them to treat at first only the amplitude fluctuations within the FRG and to then use the result as an initial condition for the phase part of the action, which they subsequently solved via the well-known Kosterlitz-Thouless flow equations. As a consequence, they recovered the usual BKT transition while still incorporating amplitude fluctuations via a rescaled initial phase stiffness. Noticeably, within their procedure amplitude fluctuations stay always gapped, in contrast to the previous FRG treatments mentioned above.

In this work we study the effect of amplitude fluctuations on the BKT transition by means of a new FRG approach, which combines the lattice FRG approach developed by Machado and Dupuis \cite{Machado10} with a dual representation of a generalized XY-model, which explicitly involves the vortex excitations via an integer-valued field. In contrast to the rather unconventional real-space RG approach developed by Kosterlitz and Thouless \cite{Kosterlitz73,Kosterlitz74}, our FRG approach is based on a straightforward application of the established FRG machinery in momentum space. We show that weak amplitude fluctuations can be taken into account as an effective interaction between the vortices. While this effective interaction leads to finite corrections of non-universal quantities like the value of the critical temperature, we find that it does not spoil the BKT transition, in agreement with general expectations.

The rest of this article is organized as follows: In Sec.~\ref{sec:dualRepresentations} we review several dual representations of the XY-model. We then use in Sec.~\ref{sec:initialCondXY} one particular dual representation as a starting point for our FRG approach, which is based on the lattice FRG formalism introduced by Machado and Dupuis \cite{Machado10}. In Sec.~\ref{sec:BKTinXYfromFRG} we solve the resulting infinite hierarchy of flow equations for the relevant and marginal couplings of the model and rederive the well-known Kosterlitz-Thouless flow equations for the XY-model within our FRG formalism. In Sec.~\ref{sec:amplitudeFlucts} we add weak amplitude fluctuations to our dual vortex action for the XY-model and show that they do not qualitatively change the BKT transition. Finally, we extend our procedure to the strongly anisotropic classical Heisenberg XXZ-model in Sec.~\ref{sec:appHeisenberg} and demonstrate that it also exhibits a BKT transition. Technical details regarding the derivation and solution of the flow equations for the XY-model are given in Appendix~\ref{sec:appTechDetails}, while in Appendix~\ref{sec:appFlowg} we derive an additional flow equation due to the amplitude fluctuations.

\section{Duality transformations of the XY-model}
\label{sec:dualRepresentations}

In this section we summarize various representations of the partition function of the two-dimensional XY-model. The mapping between these representations is constructed by means of duality transformations \cite{Jose77,Chaikin95,Herbut07}. Our dual lattice FRG approach will employ one particular representation involving integer degrees of freedom representing vortices. In this representation, the BKT transition is described by a Gaussian fixed point, which enables us to study the effect of longitudinal fluctuations on the BKT transition in two-dimensional Bose systems in a straightforward way.

The Hamiltonian of the classical XY-model
with nearest neighbor exchange interaction $J$ is
\begin{equation}
 H_{\rm XY} = - J \sum_{ i , \mu  } \bd{s}_i \cdot \bd{s}_{i + \mu } = 
 - J  \sum_{ i , \mu  } \cos ( \theta_{i+ \mu} - \theta_{i  } ),
\label{eq:hamiltonianXY}
\end{equation}
where $\bd{s}_i = \bd{e}_x \cos \theta_i + \bd{e}_y \sin \theta_i$ 
are unit vectors representing classical spins located at the
sites $\bd{r}_i$ of a square lattice with lattice spacing $a$.
The subscript $i + \mu$
represents the sites $\bd{r}_i + \bd{a}_{\mu}$,
where $\bd{a}_{\mu}$ connects nearest neighbor sites
in the direction $\mu = x,y$.
We would like to calculate the partition function 
 \begin{equation}
 Z_{\rm XY} = \prod_{i} 
 \left( \int_{ 0 }^{ 2 \pi} \frac{d \theta_i }{2 \pi } \right) e^{ \frac{1}{\tau}
 \sum_{ i , \mu  } \cos ( \Delta_{\mu} \theta_i  ) },
 \label{eq:actionXYmodel}
 \end{equation}
where we have introduced the dimensionless temperature
\begin{equation}
\tau = T / J
\end{equation}
and the lattice  derivative
 \begin{equation}
 \Delta_{\mu} \theta_i  = \theta_{i + \mu} - \theta_i.
 \label{eq:defThetaDiff}
 \end{equation}
In order to facilitate analytical treatment of the partition function we use the Villain 
approximation \cite{Villain75},
\begin{align}
e^{\frac{1}{\tau} \cos(\Delta_\mu \theta_i)} \approx R_V (1/\tau) \sum_{n_{i \mu} = -\infty}^\infty \exp \left[- \frac{(\Delta_\mu \theta_i - 2\pi n_{i \mu})^2}{2 \tau_V (1/\tau)} \right],
\label{eq:VillainApproximation}
\end{align}
which becomes exact in the limit of low as well as high temperature \cite{Janke85}.
The functions $R_V$ and $\tau_V$ are determined by expanding both sides of Eq.~\eqref{eq:VillainApproximation} in a Fourier series and demanding equality for the three lowest Fourier coefficients, 
leading to the identification \cite{Janke85}
\begin{align}
R_V (x) &= I_0 (x) \sqrt{\frac{2\pi}{\tau_V (x)}},
\label{eq:VillainR}
\\
\tau_V (x) &= 2 \ln \left( \frac{I_0 (x)}{I_1 (x)} \right),
\label{eq:VillainTau}
\end{align}
where $I_0$ and $I_1$ are modified Bessel functions of the first kind. Dropping $R_V$ as it just leads to a constant rescaling of the partition function we find
 \begin{align}
 Z_{\rm Villain} & = \prod_i 
 \left( \int_{ 0 }^{ 2 \pi} \frac{d \theta_i }{2 \pi } \right) \prod_{i, \mu} 
 \left(  \sum_{n_{i \mu} = - \infty }^{\infty}
\right) 
 \nonumber
 \\
 & \times
 \exp \left[ - \sum_{ i , \mu  } \frac{( \Delta_{\mu} \theta_i  - 2 \pi n_{i \mu })^2}{2\tau_V (1/\tau)}  \right].
 \label{eq:ZVillain}
 \end{align}
Since $\tau_V (1/\tau)$ only amounts to a rescaling of the dimensionless temperature, we will in the following write $\tau$ instead of $\tau_V (1/\tau)$ to simplify the notation. The integers $n_{i \mu}$ label the periods of
$\cos ( \Delta_{\mu} \theta_i )$.
It is convenient to eliminate the $n_{i \mu}$
in favor of another set of integers $p_{ i \mu }$ by means of the following mathematical
identity,
 \begin{equation}
 \sum_{ n = - \infty}^{\infty} e^{ -  ( x - 2 \pi n )^2 /(2 \tau)}
 = \sqrt{ \frac{\tau}{ 2 \pi } } \sum_{ p = - \infty}^{\infty} e^{ - \tau  p^2/2 - i p x },
 \label{eq:poisson}
 \end{equation}
which follows by specifying $ f ( x ) = e^{  -x^2 / ( 2 \tau )}$ in
the Poisson summation formula \cite{Henrici91}
 \begin{equation}
 \sum_{ n = - \infty}^{\infty} f ( x - 2 \pi n ) = \frac{1}{2 \pi}
 \sum_{ p = - \infty}^{\infty} e^{ - i p x } \int_{ - \infty}^{\infty} dx^{\prime} e^{ i p x^{\prime}}
 f ( x^{\prime} ).
 \label{eq:poisson1}
 \end{equation}
With the help of the identity (\ref{eq:poisson})
we obtain from Eq.~(\ref{eq:ZVillain})
 \begin{align}
 Z_{\rm Villain} &= \left( \frac{\tau}{2 \pi} \right)^N \prod_i 
 \left( \int_{ 0 }^{ 2 \pi} \frac{d \theta_i }{2 \pi } \right) \prod_{i, \mu} 
 \left(  \sum_{p_{i \mu} = - \infty }^{\infty}
\right) 
 \nonumber
 \\
 &\times
 \exp \left[ - \sum_{ i , \mu } \left( \frac{\tau}{2}
 p_{ i \mu}^2 + i p_{ i \mu} \Delta_{\mu} \theta_i  \right) \right],
 \label{eq:ZVillain2}
 \end{align}
where $N$ is the number of lattice sites. The second term in the exponent can be written as
 \begin{align}
  -i \sum_{ i , \mu} p_{ i \mu} \Delta_{\mu} \theta_i &= -i \sum_{ i , \mu} p_{ i \mu}
 ( \theta_{ i + \mu} - \theta_i )
 \nonumber
 \\
 &= i \sum_ i \theta_i \left[ \sum_{\mu}  ( p_{ i \mu } - p_{ i - \mu ,  \mu} ) \right]
 \nonumber
 \\
 &\equiv  i \sum_i \theta_i \bd{\Delta} \cdot \bd{p}_i ,
 \end{align}
 where the $\bd{p}_i = ( p_{i, x} , p_{i , y })$ are two-component vectors over the integers
representing a current-like degree of freedom
and
the lattice divergence of these currents is defined by \cite{Herbut07}
 \begin{equation}
 \bd{\Delta} \cdot \bd{p}_i = \sum_{\mu}  ( p_{ i \mu } - p_{ i - \mu ,  \mu} ).
 \label{eq:defPDelta}
 \end{equation}
Carrying out the integrations  over the angles $\theta_i$ in
Eq.~(\ref{eq:ZVillain2}) enforces local constraints on the $p_{i \mu}$ at each site, which can be written as
a vanishing lattice divergence.  The
partition function of the Villain model can then be written as a constrained sum over the
integers $p_{ i \mu}$,
 \begin{align}
 Z_{\rm Villain} &=
 \left( \frac{\tau}{2 \pi} \right)^N 
\prod_{i, \mu} \left( \sum_{p_{i\mu}=-\infty }^{\infty}
  \delta_{ \bd{\Delta} \cdot \bd{p}_i, 0 } \right)
 e^{  -   \frac{\tau}{2}  \sum_{ i , \mu }  
 p_{ i \mu}^2 }.
 \label{eq:ZVillain3}
 \end{align}
Now we can use the fact that the currents $\bd{p}_i$ 
have vanishing lattice divergence to express them in terms of a new set of integers
$m_i$ attached to the sites of the dual lattice 
\begin{figure}
\includegraphics[width=\columnwidth]{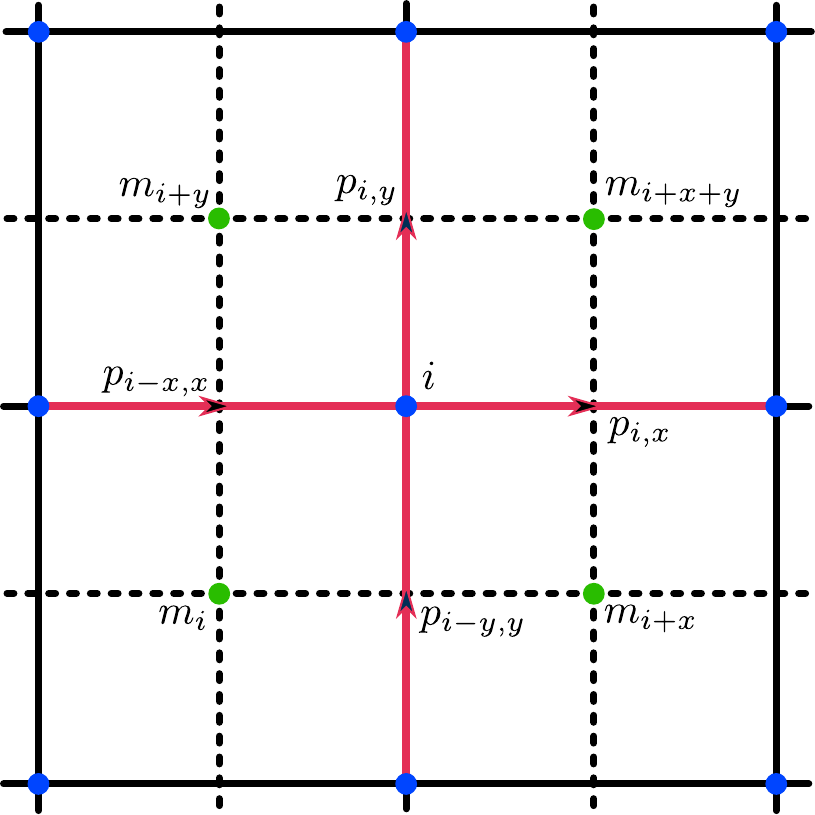}
\caption{
%(Color online)
Schematic representation of the physical lattice (blue dots), the dual lattice (green dots), and of the related fields. The red arrows denote the currents $p_{i \mu}$ flowing into and out of the site $i$ of the physical lattice in direction $\mu =x,y$. The $m$-field  introduced in Eqs.~(\ref{eq:defMfieldA}--\ref{eq:defMfieldD}) is defined on the sites of the dual lattice.
}
\label{fig:dualLattice}
\end{figure}
as follows \cite{Jose77,Savit78,Herbut07},
 \begin{subequations}
 \begin{align}
 p_{ i , x } &= m_{i +x+y} - m_{ i+x} = \Delta_y m_{i+x},
 \label{eq:defMfieldA}
 \\
 p_{i,y} &= m_{i+y} - m_{i+x+y} = - \Delta_x m_{i+y},
 \\
 p_{i-x,x} &= m_{ i+y} - m_i = \Delta_y m_i,
 \\
 p_{ i-y,y} &= m_i - m_{i+x} = - \Delta_x m_i.
 \label{eq:defMfieldD}
 \end{align}
\end{subequations}
A graphical representation of the dual lattice as well as of the
association of the currents $p_{i \mu }$ with the links
of the original lattice is shown in Fig.~\ref{fig:dualLattice}.
The four conditions (\ref{eq:defMfieldA}--\ref{eq:defMfieldD})
can be written compactly as a lattice curl,
 \begin{equation}
 p_{ i \mu} =  \epsilon_{\mu \nu} \Delta_{\nu}  m_{i+ \mu},
 \end{equation}
where the matrix elements of the  
antisymmetric $\epsilon$-tensor  are $\epsilon_{ xx } = \epsilon_{yy} =0$ and $\epsilon_{xy} = - \epsilon_{yx} = 1$.
The condition $\bd{\Delta} \cdot \bd{p}_i =0$ is then automatically satisfied
and we can represent the partition function of the Villain model as an unconstrained sum
over integer variables attached to the sites of the dual lattice,
 \begin{equation}
 Z_{\rm Villain} =  \left( \frac{\tau}{2 \pi} \right)^N 
\prod_i  \left( \sum_{ m_i = - \infty}^{\infty} \right)
 e^{-   S_{\rm dual} [ m ]  },
 \label{eq:dual1}
 \end{equation}
where the dual vortex action is the following quadratic functional of integers $m_i$ associated with the
sites $\bd{R}_i$  of the dual lattice,
 \begin{equation}
 S_{\rm dual} [ m ] =  \frac{\tau}{2} \sum_{ i, \mu} (\Delta_{\mu}  m_{ i } )^2
 = \frac{\tau}{2} \sum_{ i, \mu} ( m_{ i + \mu  } - m_i )^2.
 \label{eq:Sdualm}
 \end{equation}
This dual vortex action will be the starting point of our FRG calculation
in Sec.~\ref{sec:initialCondXY}. Note that the partition function \eqref{eq:dual1} is actually divergent, since the dual vortex action \eqref{eq:Sdualm} depends only on differences of the $m$-field. The reason for this divergence is that the relations (\ref{eq:defMfieldA}--\ref{eq:defMfieldD}) define the $m$-field only up to an overall additive constant. In principle, we should therefore fix the $m$-field at an arbitrary lattice site to some integer $m_f$ \cite{Ney95}. However, we may then sum over $m_f$ without changing the physics as this only amounts to a constant rescaling of the partition function, so that we finally obtain the simpler expression \eqref{eq:dual1}. Furthermore, in Sec.~\ref{sec:initialCondXY} we will introduce a regulator, which leads to a finite zero-momentum mode of the dispersion of the $m$-field. As a result, the divergence of the partition function is cured for any finite value of the cutoff.

For completeness, we conclude this section by establishing the relation
of the dual vortex action (\ref{eq:Sdualm}) to the so-called Coulomb gas representation of the
XY-model. Therefore we make use of the Poisson summation formula (\ref{eq:poisson1})
once more to  
eliminate the integers $m_i$ in favour of a set of real continuous 
variables $\phi_i$ and another set of integers $q_i$,
 \begin{equation}
 Z_{\rm Villain}  =   \left( \frac{\tau}{2 \pi} \right)^N 
\prod_i \left( \int_{ - \infty}^{\infty} d \phi_i 
\sum_{ q_i = - \infty}^{\infty} \right) e^{ - S_{\rm dual} [ \phi, q ] } ,
 \label{eq:dual2}
 \end{equation}
where the dual action is now given by
 \begin{equation} 
  S_{\rm dual} [ \phi, q ] = \frac{\tau}{2} \sum_{i, \mu }  (  \Delta_{\mu} \phi_{ i  } )^2
 + 2 \pi i  \sum_i q_i \phi_i.
 \end{equation}
The $q_i$ are called vortex charges; in the Coulomb gas analogy \cite{Minnhagen87},
the $q_i$ correspond to the charges of the particles in units of the elementary charge.
The easiest way to see the equivalence of Eqs.~(\ref{eq:dual1}) and (\ref{eq:dual2})
is to use the distributional form of the Poisson summation formula (\ref{eq:poisson1}),
 \begin{equation}
 \sum_{ q_i = - \infty}^{\infty} e^{ - 2 \pi i q_i \phi_i } = \sum_{ m_i = - \infty}^{\infty}
 \delta ( \phi_i - m_i ).
 \end{equation}
Integrating over the fields $\phi_i$ then replaces $\phi_i \rightarrow m_i$ in the rest of the integrand so that
we recover Eq.~(\ref{eq:Sdualm}).
However, we may also first perform the Gaussian integration in Eq.~(\ref{eq:dual2})
over the field $\phi_i$ to obtain 
\begin{align}
 Z_{\rm Villain}  &= \sqrt{N} \left( \frac{\tau}{2 \pi} \right)^N e^{ \frac{1}{2} \sum_{\bd{k} \neq 0} \ln ( 2 \pi / \omega_{\bd{k}} ) }
 \nonumber
 \\
 &\times
\prod_i \left( \sum_{ q_i = - \infty}^{\infty} \right) \delta \bigg( \sum_i q_i \bigg) e^{ -   \frac{1}{2} \sum_{ij} V_{ij} q_i q_j  },
 \label{eq:Svortex}
 \end{align}
where the delta distribution enforces the constraint of vanishing total vortex charge (the appearance of the delta distribution instead of a Kronecker delta is directly related to the divergence of the partition function \eqref{eq:dual1} as discussed above) and the interaction $V_{ij}$ is given by
 \begin{equation}
 V_{ij} = \frac{( 2 \pi)^2}{N} \sum_{\bd{k} \neq 0} \frac{ e^{ i \bd{k} \cdot ( \bd{R}_i  - \bd{R}_j )}}{ \omega_{\bd{k}} }.
 \label{eq:Vij}
 \end{equation}
Here we have defined the dimensionless dispersion
 \begin{equation}
 \omega_{\bd{k}} = 4 \tau ( 1 - \gamma_{\bd{k}} ),
 \label{eq:omegakdef}
 \end{equation}
where $ \gamma_{\bd{k}}$  is the nearest neighbor structure factor on a square lattice,
 \begin{equation}
 \gamma_{\bd{k}} = \frac{1}{2} [ \cos ( k_x a ) + \cos ( k_y a ) ].
 \end{equation}
In the thermodynamic limit, the summation in Eq.~(\ref{eq:Vij})
should be replaced by an integration over the
first Brillouin zone,
 \begin{equation}
 V_{ij} = a^2 \int_{ - \pi /a }^{ \pi / a }  d k_x
 \int_{ - \pi /a }^{ \pi / a }  d k_y  
 \frac{ e^{i \bd{k} \cdot ( \bd{R}_i - \bd{R}_j ) }}{\omega_{\bd{k}}}.
 \end{equation}
Using the fact that for small wavevectors
$ \gamma_{\bd{k}} \approx 1 -   \bd{k}^2 a^2 /4$ and hence
$\omega_{\bd{k}} \approx \tau a^2  \bd{k}^2 $, we see that
the integral is infrared divergent.
To regularize this divergence, we assume that the system has a finite length $L$
and rewrite the action in Eq.~(\ref{eq:Svortex}) as
 \begin{align}
\frac{1}{2} \sum_{ij} \tilde{V}_{ij} q_i q_j 
 &= \frac{1}{2} \sum_{ij} V_{ij} q_i q_j 
 - \frac{V_0}{2} \Bigl( \sum_i q_i \Bigr)^2
 \end{align}
with
\begin{equation}
 V_0 = \frac{( 2 \pi)^2}{N} \sum_{\bd{k} \neq 0} \frac{1}{\omega_{\bd{k}}},
\end{equation}
which is allowed due to the constraint $\sum_i q_i = 0$.
This regularized interaction
 \begin{equation}
 \tilde{V}_{ij} = V_{ij} - V_0 = 
 \frac{ (2 \pi)^2 }{N} \sum_{\bd{k} \neq 0} \frac{  e^{i    \bd{k} \cdot ( \bd{R}_i - \bd{R}_j ) } -1  }{\omega_{\bd{k}}}
 \end{equation}
then has a finite limit for $L \rightarrow \infty$ and behaves as
\begin{align}
\tilde{V}_{ij} \sim - \ln (| \bd{R}_i - \bd{R}_j | / a)
\end{align}
for large distances $ | \bd{R}_i - \bd{R}_j | \gg a $.

\section{Dual lattice FRG}
\label{sec:initialCondXY}

In the derivation of the exact hierarchy of FRG flow equations for a given many-body system, one 
usually assumes that the partition function can be expressed as a functional integral involving 
continuous fields \cite{Kopietz10}. 
However, as recently pointed out by Machado and Dupuis \cite{Machado10}, this assumption is really not necessary, so that the FRG formalism can also be applied to systems involving degrees of 
freedom which 
are  parametrized in terms of a set of integers. The basic idea of their lattice FRG is to start the flow in the local limit of decoupled lattice sites, thus retaining information about local fluctuations in the initial conditions. This is similar \cite{Machado10} to the hierarchical reference theory of fluids developed by Parola and Reatto \cite{Parola84,Parola85,Parola12}, which starts with a soluble reference system containing only strongly repulsive short-range interactions and then progressively includes long-range interactions via a RG procedure.

At the first sight it is not clear whether the dual vortex representation (\ref{eq:dual1})
or the Coulomb gas representation (\ref{eq:Svortex}) will be most suitable for
deriving the Kosterlitz-Thouless RG equations within the lattice FRG formalism
proposed by Machado and Dupuis \cite{Machado10}.
While we have explored both possibilities, in the rest of this work we shall present results only for 
the dual vortex representation (\ref{eq:dual1}), which we found to be advantageous due to the fact that the
dual action $S_{\rm dual} [ m ]$ is simply given by squares of lattice derivatives.

\subsection{Exact FRG flow equations}
\label{subsec:flow}

Starting point is the dual representation (\ref{eq:dual1})
of the partition function of the XY-model as a nested sum over integers $m_i$
representing the dual vortex degrees of freedom.
Introducing the Fourier components of the $m_i$ via
 \begin{equation}
 m_i   =  \frac{1}{\sqrt{N}} \sum_{\bd{k}} e^{ i \bd{k} \cdot \bd{R}_i }
  m_{\bd{k}},
 \end{equation}
the dual vortex action (\ref{eq:Sdualm}) can be written as
\begin{equation}
 S_{\rm dual} [ m ] =  \frac{\tau}{2} \sum_{ i, \mu} (\Delta_{\mu}  m_{ i } )^2
 = \frac{1}{2} \sum_{\bd{k}} \omega_{\bd{k}}   m_{ - \bd{k}} m_{\bd{k}},
 \label{eq:Sdualmk}
 \end{equation}
with $\omega_{\bd{k}} = 4 \tau ( 1 - \gamma_{\bd{k}})$, see Eq.~(\ref{eq:omegakdef}).
Note that the low-temperature phase of the original XY-model maps onto the high-temperature phase of the dual model. We therefore expect that for small $\tau$ the
dual model will be gapless, while for large $\tau$ the renormalized spectrum
will exhibit a gap associated with a finite screening length.

\begin{figure}
\includegraphics[width=\columnwidth]{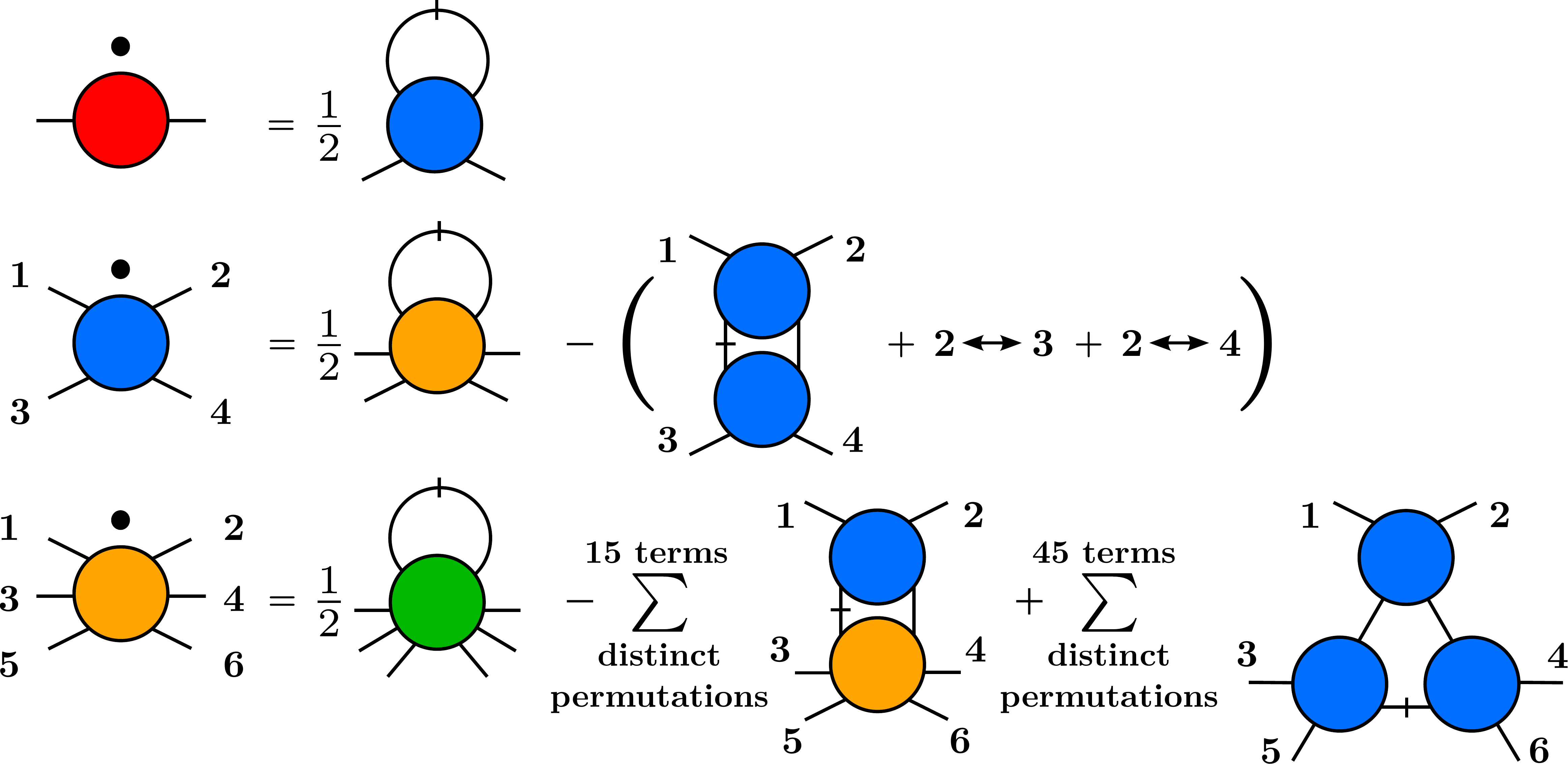}
\caption{
%(Color online)
Graphical representation of the exact FRG flow equations 
for the irreducible vertices with two and four legs, see
Eqs.~(\ref{eq:flow2}) and (\ref{eq:flow4}), as well as for the irreducible six-point vertex \cite{Kopietz01}. Solid lines represent the cutoff-dependent propagator
$G_{\lambda} ( \bd{q})$, while slashed solid lines denote the corresponding
single-scale propagator $\dot{G}_{\lambda} ( \bd{q} )$.
The dot over the vertex on the left-hand side represents the derivative
with respect to the cutoff parameter $\lambda$. We have labelled the external momenta by integers, omitting them where there is no danger of ambiguity.
}
\label{fig:flowexact}
\end{figure}

If the $m_i$ were continuous variables, the dual action (\ref{eq:Sdualmk})
would represent a Gaussian field theory which does not exhibit any phase transition.
The BKT transition must therefore be related to the discreteness
of the $m_i$. 
To derive the  Kosterlitz-Thouless 
RG equations from the dual representation (\ref{eq:dual1})
within the framework of the lattice FRG \cite{Machado10},
we introduce a bandwidth cutoff $\lambda$ 
and replace the dual action (\ref{eq:Sdualmk}) by the cutoff-dependent action
   \begin{equation}
 S_{\lambda} [ m ] = \frac{1}{2} \sum_{\bd{k}} [ \omega_{\bd{k}} + R_{\lambda} ( \bd{k} ) ]  
 m_{ - \bd{k}} m_{\bd{k}} ,
 \label{eq:Sdualreg}
 \end{equation}
where the cutoff-dependent regulator function $R_{\lambda} ( \bd{k} )$
is given by \cite{Machado10}
 \begin{equation}
 R_{\lambda} ( \bd{k} ) = ( \lambda - \omega_{\bd{k}} )
 \Theta( \lambda - \omega_{\bd{k}} ) .
 \label{eq:regulator}
\end{equation}
Identifying the initial value $\lambda_0$ of the cutoff with the total bandwidth
of the dispersion, i.e.,
 \begin{equation}
 \lambda_0 = {\rm max} \{  \omega_{\bd{k}} \} =   8 \tau,
 \end{equation}
we see that at  $\lambda = \lambda_0$  the regularized dispersion is
constant for all wavevectors  $\bd{k}$,
 \begin{equation}
 \omega_{\bd{k}} + R_{\lambda_0} ( \bd{k} ) = \lambda_0,
 \end{equation}
so that at the initial cutoff scale the dual action is local,
 \begin{align}
 S_{\lambda_0} [ m ] &= \frac{\lambda_0}{2} \sum_{\bd{k}}  
 m_{ - \bd{k}} m_{\bd{k}}  = \frac{\lambda_0}{2} \sum_i m_i^2.
 \label{eq:Sinit}
 \end{align}
In order to define the average effective action, we introduce the cutoff-dependent
generating functional $W_{\lambda} [ h ]$   of the connected correlation functions 
(Schwinger functional),
\begin{align}
 e^{ W_{\lambda} [ h ] } &= 
  \prod_i  \left(
\sum_{ m_i = - \infty}^{\infty} \right) 
e^{  - S_{\lambda} [ m ] + \sum_i  h_i m_i  },
 \label{eq:Wdef}
 \end{align} 
which is a functional of the real-valued source fields $h_i$.
The average effective action $\Gamma_\lambda [\bar{m}]$ is then
defined via the subtracted Legendre transformation,
 \begin{equation}
 \Gamma_{\lambda} [ \bar{m} ]  =  \sum_i   h_i \bar{m}_i  - W_{\lambda} [ h[\bar{m}] ] 
   - \frac{1}{2} \sum_{\bd{k}}
 R_{\lambda} ( \bd{k} ) \bar{m}_{- \bd{k}} \bar{m}_{\bd{k}},
 \label{eq:Gammadef}
 \end{equation}
where on the right-hand side 
it is understood that the sources
$h_i = h_i [ \bar{m} ]$ should be considered as functionals of the expectation values 
$\langle m_i \rangle$ by solving
\begin{align}
  \frac{ \delta W_{\lambda} [  h ]}{
 \delta h_i }  &=   \langle m_i \rangle  \equiv \bar{m}_i  .
 \label{eq:naverage}
 \end{align}
Taking derivatives of $W_{\lambda} [ h ]$ with respect to the sources,
we obtain the connected correlation functions of the dual integers $m_i$.
In particular, 
the two-point connected correlation function between $m_i$ and $m_j$  
at the dual sites $\bd{R}_i$ and $\bd{R}_j$ is
 \begin{align}
\frac{ \delta^2 W_{\lambda} [  h ]}{
 \delta h_i \delta h_j } &= 
  \langle m_i m_j \rangle  -   \langle m_i \rangle   \langle m_j \rangle ,
 \label{eq:Gcon}
 \end{align}
where for an arbitrary functional $F [ m ]$ the average symbol is defined by
 \begin{align}
\langle F [ m ] \rangle = 
 \frac{  \prod_i \left(   
\sum_{ m_i = - \infty}^{\infty} \right) 
  e^{  - S_{\lambda} [  m ] + \sum_i  h_i m_i  }  F [ m ]   }{  \prod_i \left( 
\sum_{ m_i = - \infty}^{\infty} \right) 
  e^{  - S_{\lambda} [  m ] + \sum_i  h_i  m_i }}.
 \end{align}
Following the usual steps \cite{Kopietz10}, it is now straightforward
to derive an exact hierarchy of FRG flow equations for the
irreducible vertices of our model.  
It is important to emphasize that this derivation does not require that the theory
can be defined in terms of some functional integral over continuous fields.
The average effective action  $\Gamma_{\lambda} [ \bar{m} ]$
defined in Eq.~(\ref{eq:Gammadef}) therefore
satisfies the exact FRG flow equation \cite{Wetterich93,Kopietz10}
\begin{align}
   \partial_{\lambda} \Gamma_{\lambda} [ \bar{m} ] &=
\frac{1}{2} \sum_{\bd{k}} ( \partial_{\lambda} R_{\lambda} ( \bd{k} ) )
 \bigl[
   \mathbf{\Gamma}^{\prime \prime}_{\lambda} [ \bar{m} ]
  +  \mathbf{R}_{\lambda}   \bigr]^{-1}_{ \bd{k} , - \bd{k}}      ,
 \label{eq:wetterichm}
 \end{align}
where 
 $\mathbf{\Gamma}^{\prime \prime}_{\lambda} [ \bar{m} ]$ 
and $ \mathbf{R}_{\lambda} $
are infinite matrices in the momentum labels with matrix elements given by
 \begin{align}
  \left( \mathbf{\Gamma}^{\prime \prime}_{\lambda} [ \bar{m} ]  \right)_{ \bd{k} \bd{k}^{\prime} }
 &= \frac{ \delta^2 \Gamma_{\lambda} [ \bar{m} ] }{
 \delta \bar{m}_{\bd{k} } \delta \bar{m}_{ \bd{k}^{\prime} } },
 \\
\left(  \mathbf{R}_{\lambda} \right)_{\bd{k} \bd{k}^{\prime}} &= \delta_{ \bd{k} , - \bd{k}^{\prime}}
 R_{\lambda} ( \bd{k}^{\prime} ).
 \end{align}
For $\bar{m} =0$, the second-derivative matrix of the average effective action
 $\Gamma_{\lambda} [ \bar{m} ]$ is diagonal in the momentum labels,
 \begin{equation}
  \left( \mathbf{\Gamma}^{\prime \prime}_{\lambda} [ 0 ]  \right)_{ \bd{k} \bd{k}^{\prime} }
 = \delta_{ \bd{k} , - \bd{k}^{\prime} } \Gamma^{(2)}_{\lambda} ( \bd{k}^{\prime} ),
 \end{equation}
where the Fourier transform of the connected
two-point function for vanishing sources 
is by construction related to $ \Gamma^{(2)}_{\lambda} ( \bd{k} ) $
as follows,
 \begin{align}
 G_{\lambda} ( \bd{k} ) &= \frac{1}{N} \sum_{ i j } e^{i  \bd{k} \cdot ( \bd{R}_i - \bd{R}_j ) }
 \langle m_i m_j \rangle_{ h_i  = 0  } 
 \nonumber
 \\
 &=
\bigl(
   \mathbf{\Gamma}^{\prime \prime}_{\lambda} [ 0 ]
  +  \mathbf{R}_{\lambda}   \bigr)^{-1}_{ \bd{k} , - \bd{k}}  =
 \left[ \Gamma^{(2)}_{\lambda} ( \bd{k} ) + R_{\lambda} ( \bd{k} ) \right]^{-1}.
 \label{eq:propagator}
 \end{align}
The expansion of $\Gamma_{\lambda} [ \bar{m} ]$ 
in powers of the Fourier components  $\bar{m}_{\bd{k}}$
of the expectation values $\bar{m}_i$ 
defines the irreducible vertices
\footnote{In Eq.~\eqref{eq:Gammadef} we have subtracted the cutoff term, so that the regularized two-point vertex $\Gamma^{(2)}_{\lambda} ( \bd{k} ) + R_{\lambda} ( \bd{k} )$ can be identified with the inverse propagator, see Eq.~\eqref{eq:propagator}. This differs slightly from the convention in Ref.~[\onlinecite{Kopietz10}] where, using a different subtraction, $\Gamma^{(2)}_{\lambda} ( \bd{k} )$ can be identified with the self-energy (i.e., the irreducible two-point vertex).}.
By symmetry the expansion involves only even powers,
\begin{align}
  \Gamma_{\lambda} [  \bar{m} ] &= \Gamma^{(0)}_{\lambda}
 + \frac{1}{2} \sum_{\bd{k}} 
 \Gamma^{(2)}_{\lambda} ( \bd{k} )
 \bar{m}_{- \bd{k}} \bar{m}_{\bd{k}} 
 \nonumber
\\
 &+
 \sum_{n =2}^{\infty} \frac{1}{(2n)!N^{n-1}} \sum_{ \bd{k}_1 \ldots \bd{k}_{2n} }
 \delta_{ \bd{k}_1 + \ldots + \bd{k}_{2n},0 }
 \nonumber
 \\
 &\hspace{10mm}  \times \Gamma^{(2n)}_{\Lambda} ( \bd{k}_1,  \ldots , \bd{k}_{2n} )    \bar{m}_{\bd{k}_1} \ldots \bar{m}_{\bd{k}_{2n}}
 \nonumber
 \\  
 &= 
\Gamma^{(0)}_{\lambda}
 +  \frac{1}{2} \sum_{\bd{k}} 
 \Gamma^{(2)}_{\lambda} ( \bd{k} )
 \bar{m}_{- \bd{k}} \bar{m}_{\bd{k}} 
 \nonumber
 \\
 &+ \frac{1}{ 4! N } \sum_{ \bd{k}_1 \ldots \bd{k}_4 }
 \delta_{ \bd{k}_1 + \bd{k}_2 + \bd{k}_3 + \bd{k}_4 , 0 }  \Gamma^{ (4)}_{\lambda} ( \bd{k}_1 , \bd{k}_2 , \bd{k}_3 , \bd{k}_4  )
 \nonumber
 \\
 &\hspace{20mm} \times
 \bar{m}_{\bd{k}_1 } \bar{m}_{\bd{k}_2 }
   \bar{m}_{\bd{k}_3 } \bar{m}_{\bd{k}_4 } 
  +   \ldots,
 \label{eq:vertexpm}
 \end{align}
where the ellipsis denotes terms involving six and higher powers of
$\bar{m}$. Setting $\bar{m} = 0$ in Eq.~(\ref{eq:wetterichm})
we obtain the exact flow of the free energy, 
  \begin{align}
 \partial_{\lambda} \Gamma^{(0)}_{\lambda}  =  \frac{1}{2} \sum_{\bd{k}}
 G_{\lambda} ( \bd{k} )  \partial_{\lambda} R_{\lambda} ( \bd{k} )   .
 \label{eq:flow0}
 \end{align}
By expanding both sides of  Eq.~(\ref{eq:wetterichm}) to quadratic order in $\bar{m}$,
we obtain an exact FRG flow equation for the two-point vertex \cite{Kopietz10},
 \begin{align}
 \partial_{\lambda} \Gamma^{(2)}_{\lambda} ( \bd{k} ) & =
  \frac{1}{2N} \sum_{\bd{q}}
 \dot{G}_{\lambda} ( \bd{q} )
 \Gamma^{(4)}_{\lambda}  ( \bd{\bd{k}} , -  \bd{k} ,  \bd{q} ,  - \bd{q} ),
 \label{eq:flow2}
 \end{align} 
where we have introduced the single-scale propagator
 \begin{equation}
 \dot{G}_{\lambda} ( \bd{k} ) = -  
 [ G_{\lambda} (  \bd{k} ) ]^2    \partial_{\lambda} R_{\lambda} ( \bd{k} )   .
 \label{eq:singlescale}
 \end{equation}
The exact flow equation of the
four-point vertex appearing on the right-hand side of 
Eq.~(\ref{eq:flow2}) is \cite{Kopietz10}
\begin{align}
&\partial_{\lambda} \Gamma^{(4)}_{\lambda} ( \bd{k}_1 , \bd{k}_2 , \bd{k}_3, \bd{k}_4 )
\nonumber
\\
&= \frac{1}{2N} \sum_{\bd{q}} \dot{G}_{\lambda} ( \bd{q} )\Gamma^{(6)}_{\lambda}  ( \bd{k}_1 ,  \bd{k}_2, \bd{k}_3 , \bd{k}_4, \bd{\bd{q}} , -  \bd{q})
\nonumber
\\
&- \frac{1}{N} \sum_{\bd{q}} \Bigl[ \dot{G}_{\lambda} ( \bd{q} ) \Gamma^{(4)}_{\lambda} ( \bd{k}_1 , \bd{k}_2 , \bd{q} , - \bd{q} - \bd{k}_1 - \bd{k}_2 )
\nonumber
\\
&\hspace{10mm} \times G_{\lambda} ( - \bd{q} - \bd{k}_1 - \bd{k}_2 ) \Gamma^{(4)}_{\lambda} ( \bd{q} + \bd{k}_1  +  \bd{k}_2 , - \bd{q} ,  \bd{k}_3  , \bd{k}_4 )
\nonumber
\\
&\hspace{10mm} + ( \bd{k}_2 \leftrightarrow \bd{k}_3 ) + ( \bd{k}_2 \leftrightarrow \bd{k}_4 )\Bigr].
\label{eq:flow4}
\end{align}
Graphical representations of Eqs.~(\ref{eq:flow2}) and (\ref{eq:flow4}) as well as of the exact flow equation for the six-point vertex \cite{Kopietz01} are shown in Fig.~\ref{fig:flowexact}.

\subsection{Initial conditions}
\label{subsec:initialConditions}

In the lattice FRG we start the RG flow at scale $ \lambda_0 = 8 \tau = 8 T / J $
with a non-trivial initial condition.
Noting that $S_{\lambda_0} [ m ]$ in Eq.~(\ref{eq:Sinit}) is local,
we obtain from Eq.~(\ref{eq:Wdef})
 \begin{align}
  W_{\lambda_0} [ h ]  &= \sum_{i} \ln \left[
 \sum_{ m_i = - \infty}^{\infty} e^{ - \frac{\lambda_0}{2} m_i^2  + h_i m_i } \right]
 \nonumber
 \\
 &= \sum_{i} \ln \vartheta_3 \left( i h_i /2  , e^{ -\lambda_0 /2}  \right),
 \label{eq:mSumTheta}
 \end{align}
where the theta function $\vartheta_3 ( z, q )$ 
is for $| q | < 1$ defined by \cite{Abramowitz72}
 \begin{equation}
 \vartheta_3 ( z, q ) = \sum_{ m = - \infty}^{\infty} q^{m^2} e^{ 2 i m  z }
 = 1 + 2 \sum_{ m =1}^{\infty} q^{m ^2} \cos ( 2 m  z ).
 \end{equation}
Using Jacobi's identity \cite{Henrici91}
 \begin{equation}
  \vartheta_3 ( z, e^{ - \pi x } ) = \frac{1}{\sqrt{x}} e^{ - z^2 /( \pi x)  }
 \vartheta_3 \left( \frac{ z}{ i x} , e^{ - \pi /x } \right),
 \end{equation}
we may alternatively write
  \begin{align}
 W_{\lambda_0} [h]  =   \sum_i \left[ \frac{  h_i^2}{2 \lambda_0} 
 + \ln \sqrt{ \frac{2 \pi }{ \lambda_0 }}+
  \ln \vartheta_3 \left( \frac{\pi  h_i }{\lambda_0}, e^{ - 2 \pi^2 / \lambda_0 } \right)
 \right].
 \label{eq:wim}
 \end{align}
This expression is particularly useful at low temperatures
where the parameter
 \begin{equation}
 y_0 = e^{- 2 \pi^2 / \lambda_0 } = e^{ -  \pi^2 J /(4 T )},
 \label{eq:y0def}
\end{equation}
which can be interpreted as the vortex fugacity at scale $\lambda_0$,
is exponentially small so that we may expand
 \begin{equation}
   \vartheta_3 \left( \frac{\pi  h_i }{\lambda_0}, y_0 \right)
 = 1 + 2 y_0 \cos \left( \frac{2 \pi h_i }{ \lambda_0} \right)  + {\cal{O}}( y_0^4 ).
 \label{eq:thetaApproxFirst}
 \end{equation}
Then we obtain
 \begin{equation}
 \bar{m}_i = \frac{ \delta W_{\lambda_0} [ h ]}{\delta h_i } = \frac{h_i}{\lambda_0} - \frac{4 \pi y_0}{\lambda_0}
 \sin \left( \frac{ 2 \pi h_i }{\lambda_0} \right) + {\cal{O}} ( y_0^2 ),
 \end{equation}
and hence
 \begin{equation}
 h_i = \lambda_0 \bar{m}_i + 4 \pi y_0 \sin ( 2 \pi \bar{m}_i ) + {\cal{O}} ( y_0^2 ).
 \end{equation} 
The resulting initial form of the average effective action is 
 \begin{align}
 \Gamma_{\lambda_0} [ \bar{m} ] &= - N \ln \sqrt{ \frac{2 \pi }{ \lambda_0} } +
  \frac{1}{2} \sum_{\bd{k}} \omega_{\bd{k}} \bar{m}_{- \bd{k}} \bar{m}_{\bd{k}} 
 \nonumber
 \\
 &- 2 y_0 \sum_i \cos ( 2 \pi \bar{m}_i )
  + {\cal{O}} ( y_0^2 ).
 \label{eq:Gammainit}
 \end{align}
In the long-wavelength limit where
 $\omega_{\bd{k}} \approx \tau  \bd{k}^2 a^2$,
the initial average effective action (\ref{eq:Gammainit})
formally resembles the action of the sine-Gordon field theory \cite{Herbut07}.
Here, however, $\bar{m}_i = \langle m_i \rangle$ is not a quantum field 
which is integrated over,  but the expectation value
of an integer degree of freedom $m_i$  representing the vortices in the dual
action  (\ref{eq:Sdualm}).

Note that in the above derivation of Eq. \eqref{eq:Gammainit} we have expanded $W_{\lambda_0}$ in the vortex fugacity $y_0$ in order to invert the relation
\begin{align}
\mb_i = \frac{\delta W_{\lambda_0} [h]}{\delta h_i}
\label{eq:m_implicit_def}
\end{align}
analytically. The advantage of this approach is that it allows us to rederive the Kosterlitz-Thouless RG equations (Sec.~\ref{sec:BKTinXYfromFRG}) and to analytically assess the effect of amplitude (Sec.~\ref{sec:amplitudeFlucts}) and out-of-plane fluctuations (Sec.~\ref{sec:appHeisenberg}) on the BKT transition. However, in order to investigate the full phase diagram of our dual vortex model and especially the high-temperature phase where vortices proliferate, another strategy would be necessary. A possible strategy is to invert the relation \eqref{eq:m_implicit_def} numerically and to proceed within the derivative expansion.

\section{Kosterlitz-Thouless RG equations from the FRG}
\label{sec:BKTinXYfromFRG}

\subsection{Relevant and marginal couplings}

Given the fact that the FRG flow equations above
are formally exact, it should be possible to derive from these equations the celebrated
RG equations for the renormalized temperature and for the vortex fugacity 
of the XY-model first obtained by Kosterlitz and Thouless \cite{Kosterlitz73}.
Note that these authors used an unconventional real-space implementation of the
RG procedure.  
An alternative derivation of the Kosterlitz-Thouless RG equations is based on 
the mapping to the sine-Gordon model and the subsequent application of 
a real-space RG procedure to this model \cite{Herbut07}. However, as emphasized by 
Herbut \cite{Herbut07}, with this procedure one encounters an infrared divergence
which has to be regularized.
 Although physical arguments suggest a natural regularization,
it would certainly be more satisfactory to have a self-contained RG procedure which 
can be  applied automatically.
In this section we show that,  within the established machinery of the FRG,
the Kosterlitz-Thouless  RG equations can be obtained in a straightforward
way without invoking physical arguments which lie outside the framework of the FRG.

To derive the Kosterlitz-Thouless  RG within our dual lattice FRG approach,
it is sufficient to approximate the bare dispersion $\omega_{\bd{k}} $ 
by the leading term in the expansion for small wavevectors,
 \begin{align}
 \omega_{\bd{k}} &=  c_0 \bd{k}^2 + {\cal{O}} ( \bd{k}^4 ) , \quad c_0 = \tau a^2.
 \end{align} 
It is convenient to express the bandwidth cutoff $\lambda$ 
introduced in Sec.~\ref{subsec:flow} in terms of a momentum cutoff $\Lambda$ 
by setting $\lambda = c_0 \Lambda^2$ and considering all vertices as functions
of  $\Lambda$.
Moreover, to better describe the scaling in the vicinity of the BKT transition
it is convenient to multiply the regulator given in Eq.~(\ref{eq:regulator})
by an appropriate  wave function renormalization factor $c_{\Lambda} / c_0$.
Our modified regulator is therefore
 \begin{equation}
 R_{\Lambda} ( \bd{k} ) = c_{\Lambda} ( \Lambda^2 - \bd{k}^2 )
 \Theta ( \Lambda^2 - \bd{k}^2 ),
 \label{eq:RLambda}
 \end{equation}
where the scale-dependent coupling  $c_{\Lambda}$ is defined via the
long-wavelength expansion of the flowing two-point vertex,
 \begin{equation}
 \Gamma^{(2)}_{\Lambda} ( \bd{k} )  = r_{\Lambda} + c_{\Lambda} \bd{k}^2
 +  {\cal{O}} ( \bd{k}^4 ).
 \label{eq:Gamma2long}
 \end{equation} 
The reason for introducing the factor $c_{\Lambda}$ in Eq.~(\ref{eq:RLambda})
is that it simplifies  the scale-dependent propagator,
 \begin{align}
 G_{\Lambda} ( \bd{k} ) &= 
\frac{1}{ \Gamma^{(2)}_{\Lambda} ( \bd{k} ) + R_{\Lambda} ( \bd{k} )}
 \nonumber
 \\
 &= \left\{ 
\begin{array}{cc}
 ( r_{\Lambda} + c_{\Lambda} \Lambda^2 )^{-1},
 & \quad | \bd{k} | < \Lambda, 
 \\
 ( r_{\Lambda} + c_{\Lambda} \bd{k}^2 )^{-1},
 & \quad | \bd{k} | > \Lambda.
 \end{array}
 \right.
 \end{align}
For the corresponding single-scale propagator we obtain from Eq.~\eqref{eq:singlescale}
\begin{equation}
 \dot{G}_{\Lambda} ( \bd{k} ) = - \frac{ [2 c_{\Lambda} \Lambda + (\partial_{\Lambda} c_{\Lambda}) ( \Lambda^2 - \bd{k}^2 ) ]  \Theta ( \Lambda^2 - \bd{k}^2 )}{ ( r_{\Lambda}
 + c_{\Lambda} \Lambda^2 )^2 }.
 \label{eq:singlescaleLambda}
 \end{equation}
The exact FRG flow equations describing the evolution of the irreducible
vertices as we change the momentum cutoff $\Lambda$ 
can be obtained from the exact FRG flow equations with bandwidth cutoff $\lambda$
given in Sec.~\ref{subsec:flow}
by  simply replacing $\lambda \rightarrow \Lambda$.
 In particular, from the
flow equation (\ref{eq:flow2}) for the two-point vertex we obtain
the exact FRG flow equations for the couplings $r_{\Lambda}$ and
$c_{\Lambda}$ defined via the long-wavelength expansion (\ref{eq:Gamma2long}),
\begin{align}
 \partial_{\Lambda} r_{\Lambda} &=
  \frac{1}{2N} \sum_{\bd{q}}
 \dot{G}_{\Lambda} ( \bd{q} )
 \Gamma^{(4)}_{\Lambda}  (0,0, \bd{q} , -  \bd{q}  ),
 \label{eq:flowr}
 \\
\partial_{\Lambda} c_{\Lambda} &=
  \frac{1}{4N} \sum_{\bd{q} }
 \dot{G}_{\Lambda} ( \bd{q} )
 \lim_{ k \rightarrow 0} \partial_k^2
 \Gamma^{(4)}_{\Lambda}  (     \bd{k} ,  - \bd{k} ,  \bd{q} , -  \bd{q}    ),
 \label{eq:cexactflow}
 \end{align} 
where $\partial_k = \partial / \partial k$.
By simple power counting, we see that $r_{\Lambda}$ scales as $\Lambda^{-2}$
and is therefore relevant at the Gaussian fixed point manifold.
Moreover, the canonical dimension of the dimensionless coupling $c_{\Lambda}$ vanishes
so that this coupling is marginal at the Gaussian fixed point manifold. However, in two dimensions
we have to keep track of an infinite set of relevant couplings 
given by the momentum-independent parts of the irreducible vertices with $2n$ external legs,
 \begin{equation}
 u_{\Lambda}^{(2n)} = \Gamma^{(2n)}_{\Lambda} ( 0, \ldots , 0 ), \quad
 n \in \mathbb{Z}^+,
 \end{equation}
where $\mathbb{Z}^+ = \{1,2,3, \ldots\}$ denotes the set of positive integers. In two dimensions,
all couplings $u_{\Lambda}^{(2n)}$  are relevant with canonical dimension $+2$, which means that they grow as
$\Lambda^{-2}$ for $\Lambda \rightarrow 0$.
In order to recover the Kosterlitz-Thouless flow equations 
within the framework of the FRG, we have to find a way to
keep track of the flow of all $u_\Lambda^{(2n)}$. 
It turns out that we also have to
keep track of infinitely many marginal couplings $c_{\Lambda}^{(2n)}$
defined by
 \begin{equation}
 c^{(2n)}_{\Lambda}  = a^{-2} \lim_{k \rightarrow 0} \partial_k^2 
 \Gamma^{(2n)}_{\Lambda}  (     \bd{k} ,  - \bd{k} ,  0, \ldots , 0   ), \quad n \in \mathbb{Z}^+.
 \label{eq:cndef}
 \end{equation}
The above definitions of $u_{\Lambda}^{(2n)}$ and $c_{\Lambda}^{(2n)}$
parametrize the leading two coefficients in the 
long-wavelength expansion of the $2n$-point vertices with only two non-zero external momenta,
 \begin{equation}
 \Gamma^{(2n)}_{\Lambda} ( \bd{k} , - \bd{k} , 0 , \ldots , 0 ) =
 u_{\Lambda}^{(2n)} +  \frac{ k^2 a^2 }{2} c^{(2n)}_{\Lambda}    + {\cal{O}} ( k^4 ).
 \label{eq:vertexexp}
 \end{equation}
Note that with this notation
$r_{\Lambda} = u^{(2)}_{\Lambda}$ and $c_{\Lambda} = c^{(2)}_{\Lambda} a^2 /2$.

The crucial point is now that
in the vicinity of the BKT fixed point manifold,
we can explicitly solve the infinite hierarchy of RG flow equations
for all couplings $u_{\Lambda}^{(2n)}$ and $c_{\Lambda}^{(2n)}$. 
Before showing how this is possible and how the usual flow equations derived
by Kosterlitz and Thouless can be recovered from the solution,
let us discuss the initial conditions for the FRG flow
assuming that at
the initial scale $\Lambda_0$ the parameter $y_0 = e^{ - \pi^2 /( 4 \tau )}$
introduced in Eq.~(\ref{eq:y0def}) is small compared with unity.
In this regime, the proper initial condition for the
average effective action $\Gamma_{\Lambda_0} [ \bar{m} ]$
is
determined by the action (\ref{eq:Gammainit}) at the
initial value of the bandwidth cutoff $\lambda_0 = 8 \tau$,
\begin{align}
 \Gamma_{\Lambda_0} [ \bar{m} ] &= - N \ln \sqrt{ \frac{ 2 \pi}{8 \tau } }  +
  \frac{c_0}{2} \sum_{\bd{k}}  \bd{k}^2  \bar{m}_{- \bd{k}} \bar{m}_{\bd{k}} 
 \nonumber
 \\
 &\indent - 2 y_0 \sum_i \cos ( 2 \pi \bar{m}_i )
  + {\cal{O}} ( y_0^2 )
 \nonumber
 \\
 &= \Gamma_{0}^{(0)} + \frac{1}{2} 
 \sum_{\bd{k}} ( r_0 + c_0 \bd{k}^2 ) \bar{m}_{- \bd{k}} \bar{m}_{\bd{k}} 
 \nonumber
 \\
 &\indent + \sum_{ n=2}^{\infty} \frac{ u_{0}^{(2n)} }{ (2n) !}  \sum_i \bar{m}_i^{2n} + {\cal{O}} 
( y_0^2 ),
 \label{eq:Gammainit2}
 \end{align}
where
 \begin{subequations}
 \begin{align}
 \Gamma_0^{(0)} &=  - N \left[ \ln \sqrt{  \frac{ 2 \pi}{8 \tau } } +  
 2  y_0 \right],
 \\ 
 r_0  &=   ( 2 \pi )^2 2 y_0 ,
 \label{eq:r0init}
 \\
 c_0 &= \tau a^2,
 \\
 u_0^{(2n)} &= (-1)^{n+1} (2 \pi )^{2n} 2 y_0.
 \end{align}
 \end{subequations}

To derive a closed hierarchy of RG flow equations for all the relevant couplings
$u_{\Lambda}^{(2n)}$ and the marginal couplings $c_{\Lambda}^{(2n)}$,
it is instructive to consider first the  exact FRG flow equations for
the momentum-dependent vertices with two, four, and six external legs \cite{Kopietz10}
for the special case where all but two external momenta vanish. To exhibit the general structure, let us write down again 
the exact flow equation for the two-point vertex [see Eq.~(\ref{eq:flow2})],
 \begin{align}
 \partial_{\Lambda} \Gamma^{(2)}_{\Lambda} ( \bd{k} ) &=
  \frac{1}{2N} \sum_{\bd{q}}
 \dot{G}_{\Lambda} ( \bd{q} )
 \Gamma^{(4)}_{\Lambda}  ( \bd{\bd{k}} , -  \bd{k} ,  \bd{q} ,  - \bd{q} ).
 \label{eq:flow2b}
 \end{align} 
We will show later that for $n \geq 2$ all relevant couplings $u^{(2n)}_\Lambda$ are proportional to the scale-dependent fugacity $y_\Lambda$, while the marginal couplings $c^{(2n)}_\Lambda$ are only proportional to $y_\Lambda^2$. Since we assume $y_0$ and $y_\Lambda$ to be small, we can neglect the $\bd{q}$-dependence
of the four-point vertex on the right-hand side of Eq.~(\ref{eq:flow2b}) to leading order in the fugacity $y_\Lambda$.
Then the flow of the two-point vertex is given by
 \begin{align}
 \partial_{\Lambda} \Gamma^{(2)}_{\Lambda} ( \bd{k} ) &=
  \frac{1}{2N} \sum_{\bd{q}}
 \dot{G}_{\Lambda} ( \bd{q} )
 \Gamma^{(4)}_{\Lambda}  ( \bd{\bd{k}} ),
 \label{eq:flow2c}
 \end{align} 
where we have defined
 \begin{equation}
  \Gamma^{(2n)}_{\Lambda}  ({\bd{k}} ) =
 \Gamma^{(2n)}_{\Lambda}  ( {\bd{k}} , -  \bd{k} ,  0 ,   \ldots , 0 ), \quad n \in \mathbb{Z}^+.
 \end{equation}

\begin{figure}
\includegraphics[width=\columnwidth]{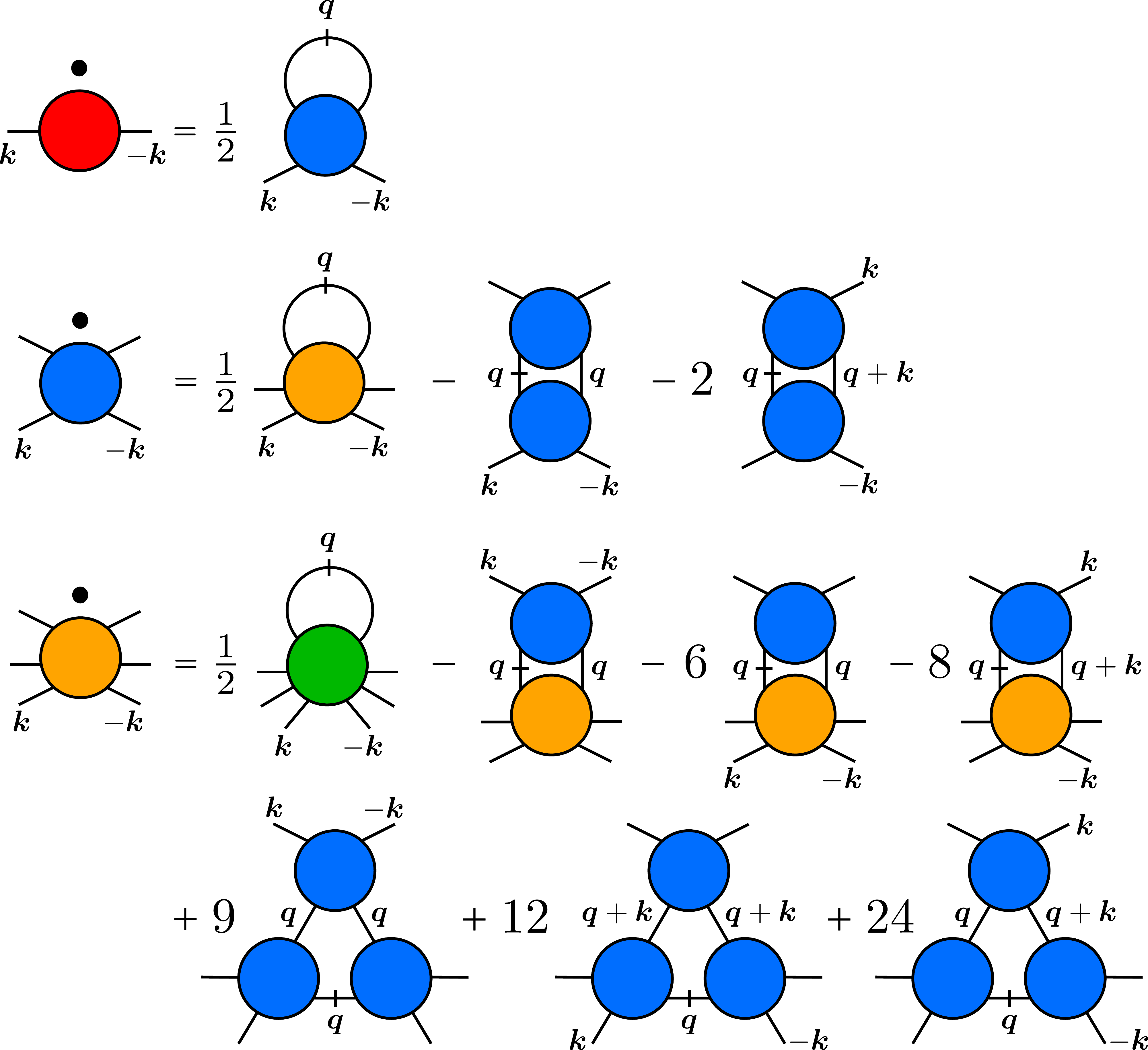}
\caption{
%(Color online)
Graphical representation of the approximate flow equations for the irreducible vertices $\Gamma^{(2)}_\Lambda (\bd{k})$, $\Gamma^{(4)}_\Lambda (\bd{k})$, and $\Gamma^{(6)}_\Lambda (\bd{k})$ as given in Eqs.~\eqref{eq:flow2c}, \eqref{eq:flow4b}, and \eqref{eq:flow6b}. Here external legs 
without labels carry vanishing momentum.
}
\label{fig:flowApprox}
\end{figure}

The  exact FRG flow equation of the effective interaction
 $\Gamma^{(4)}_{\Lambda}  ( \bd{\bd{k}} , -  \bd{k} ,  0 ,  0 )$
can be obtained
from the flow equation (\ref{eq:flow4}) by replacing $\lambda \rightarrow \Lambda$ 
and specifying the external momenta appropriately.
Analogous to Eq.~(\ref{eq:flow2c}) we can neglect the loop momentum $\bd{q}$ 
in the vertices on the right-hand side of the flow equation and obtain
 \begin{align}
 \partial_{\Lambda}    \Gamma^{(4)}_{\Lambda}  (  \bd{k} )
  &=
 \frac{1}{2N} \sum_{\bd{q}}
 \dot{G}_{\Lambda} ( \bd{q} )\Gamma^{(6)}_{\Lambda}  (  \bd{k} )    
 \nonumber
 \\
 &- \frac{1}{N} \sum_{\bd{q}} 
\dot{G}_{\Lambda} ( \bd{q} ) G_{\Lambda} ( \bd{q} )
 \Gamma^{(4)}_{\Lambda} ( \bd{k} ) \Gamma^{(4)}_{\Lambda} ( 0 )
 \nonumber
 \\
 &- \frac{2}{N} \sum_{\bd{q}} 
\dot{G}_{\Lambda} ( \bd{q} ) G_{\Lambda} ( \bd{k} + \bd{q} )
 [ \Gamma^{(4)}_{\Lambda} ( \bd{k} ) ]^2 .
 \label{eq:flow4b}
 \end{align}
Finally, within the same approximations we obtain from
the exact flow equation
for the irreducible six-point vertex shown in Fig.~\ref{fig:flowexact}
\begin{align}
&\partial_{\Lambda}    \Gamma^{(6)}_{\Lambda}  ( \bd{k} ) =  \frac{1}{2N} \sum_{\bd{q}} \dot{G}_{\Lambda} ( \bd{q} )\Gamma^{(8)}_{\Lambda}  ( \bd{k} )
\nonumber
\\
&- \frac{1}{N} \sum_{\bd{q}} \dot{G}_\Lambda (\bd{q}) G_\Lambda (\bd{q}) \Gamma^{(4)}_\Lambda (\bd{k}) \Gamma^{(6)}_\Lambda (0)
\nonumber
\\
&- \frac{6}{N} \sum_{\bd{q}} \dot{G}_\Lambda (\bd{q}) G_\Lambda (\bd{q}) \Gamma^{(4)}_\Lambda (0) \Gamma^{(6)}_\Lambda (\bd{k})
\nonumber
\\
&- \frac{8}{N} \sum_{\bd{q}} \dot{G}_\Lambda (\bd{q}) G_\Lambda (\bd{q}+\bd{k}) \Gamma^{(4)}_\Lambda (\bd{k}) \Gamma^{(6)}_\Lambda (\bd{k})
\nonumber
\\
&+ \frac{9}{N} \sum_{\bd{q}} \dot{G}_\Lambda (\bd{q}) \left[ G_\Lambda (\bd{q}) \right]^2 \Gamma^{(4)}_\Lambda (\bd{k}) \left[ \Gamma^{(4)}_\Lambda (0) \right]^2
\nonumber
\\
&+ \frac{12}{N} \sum_{\bd{q}} \dot{G}_\Lambda (\bd{q}) \left[ G_\Lambda (\bd{q}+\bd{k}) \right]^2 \left[ \Gamma^{(4)}_\Lambda (\bd{k}) \right]^2 \Gamma^{(4)}_\Lambda (0)
\nonumber
\\
&+ \frac{24}{N} \sum_{\bd{q}} \dot{G}_\Lambda (\bd{q}) G_\Lambda (\bd{q}) G_\Lambda (\bd{q}+\bd{k}) \left[ \Gamma^{(4)}_\Lambda (\bd{k}) \right]^2 \Gamma^{(4)}_\Lambda (0).
\label{eq:flow6b}
\end{align}
A graphical representation of these flow equations is shown in Fig.~\ref{fig:flowApprox}.

\subsection{Flow equation for the fugacity}

Suppose now that the momentum-independent parts
 $u_{\Lambda}^{(2n)}$ of all vertices are proportional to the same small
parameter $y_{\Lambda}$, which we arbitrarily define via
 \begin{equation}
 \Gamma^{(2)}_{\Lambda} (0 )  = u_{\Lambda}^{(2)} = r_\Lambda = (2 \pi )^2 2 y_{\Lambda}.
 \label{eq:yinit}
 \end{equation}
This assumption is certainly satisfied at the initial scale $\Lambda_0$ if we identify $y_{\Lambda_0}$ with the parameter $y_0$ in Eq.~(\ref{eq:r0init}).
The important insight is now that to linear order in $y_{\Lambda}$,
it is sufficient to retain only the first term in the
above hierarchy of flow equations involving a single vertex with two additional external legs.
Moreover, retaining only the relevant (momentum-independent) parts of the vertices which is justified for small $y_\Lambda$,
we obtain the hierarchy
 \begin{equation}
 \partial_{\Lambda} u_{\Lambda}^{(2n)} = \frac{A_{\Lambda}}{2} u_{\Lambda}^{(2n+2)}, \quad n \in \mathbb{Z}^+,
 \end{equation}
where the coefficient $A_{\Lambda}$ is within our cutoff scheme given by
\begin{align}
  A_{\Lambda} &\equiv \frac{ 1}{  N} \sum_{\bd{q}}
 \dot{G}_{\Lambda} ( \bd{q} ) =
 - \frac{ a^2 \left( \Lambda^3 c_{\Lambda} + \frac{1}{4} \Lambda^4 \partial_\Lambda c_\Lambda  \right)}{2\pi ( r_{\Lambda} + c_{\Lambda} \Lambda^2 )^2 }
 \nonumber
 \\
 &=
% \nonumber
% \\
% & \approx &  
%-  \frac{ \tau_{\Lambda}}{2 \pi}  \frac{ a^4}{ c_{\Lambda}^2 \Lambda}
  - \frac{a^2 \left( 1 + \frac{1}{4} \Lambda \frac{\partial_\Lambda c_\Lambda}{c_\Lambda} \right)}{2 \pi c_{\Lambda} \Lambda [ 1 + r_{\Lambda} / ( c_{\Lambda} \Lambda^2 )]^2} 
 \nonumber
 \\
 &=  - \frac{a^2 \left[ 1 + \mathcal{O} (y_\Lambda^2) \right]}{2 \pi c_{\Lambda}  \Lambda [
 1 + {\cal{O}} ( y_\Lambda ) ]^2}  
 =  - \frac{1}{2 \pi \tau_{\Lambda}  \Lambda } 
 + {\cal{O}} ( y_\Lambda ).
 \label{eq:Adef}
 \end{align}
Here the scale-dependent dimensionless temperature $\tau_\Lambda$ is defined as
\begin{equation}
\tau_\Lambda = \frac{c_\Lambda}{a^2}
\end{equation}
and we have used that $\partial_\Lambda c_\Lambda \propto y_\Lambda^2$, which we shall show explicitly in Sec.~\ref{sec:xyFlowMarginal}.
Introducing the logarithmic scale derivative $\partial_l = - \Lambda \partial_{\Lambda}$,
we obtain the infinite hierarchy of flow equations
 \begin{equation}
  \partial_{l} u_{\Lambda}^{(2n)}  =  
 \frac{u_{\Lambda}^{(2n+2)}}{ 4 \pi \tau_\Lambda} +
 {\cal{O}} ( y_{\Lambda}^2 ).
 \label{eq:flowuna}
 \end{equation}
To construct a solution to this hierarchy of flow equations let us make the ansatz $ u_{\Lambda}^{(2n+2)}  = b  u_{\Lambda}^{(2n)}$.
The constant $b$ can be uniquely fixed by requiring that this ansatz
is compatible with the initial condition  (\ref{eq:Gammainit2}), implying
$b = - ( 2 \pi )^2$ and hence
\begin{equation}
  u_{\Lambda}^{(2n+2)}  =   -  ( 2 \pi )^2  u_{\Lambda}^{(2n)}.
 \label{eq:urec}
 \end{equation}
All couplings $u^{(2n)}_{\Lambda}$ therefore satisfy the same flow equation
 \begin{equation}
  \partial_{l} u_{\Lambda}^{(2n)}  =  
   - \frac{\pi}{    \tau_{\Lambda}}  u_{\Lambda}^{(2n)}  +
 {\cal{O}} ( y_{\Lambda}^2 ).
 \label{eq:flowun}
 \end{equation}
Iterating the recursion (\ref{eq:urec}) we obtain with
Eq.~(\ref{eq:yinit})
 \begin{equation}
 u_{\Lambda}^{(2n)} = (-1)^{n+1} (  2 \pi )^{2n} 2 y_{\Lambda}, \quad n \in \mathbb{Z}^+.
 \label{eq:undef}
 \end{equation}
To first order in $y_{\Lambda}$,
the local part of the average effective action is therefore
\begin{align}
\Gamma^{(0)}_\Lambda + \sum_{n=1}^\infty \frac{u^{(2n)}_\Lambda}{(2n)!} \sum_i \bar{m}^{2n}_i
 = \Gamma^{(0)}_\Lambda
- 2 y_{\Lambda} \cos ( 2 \pi \bar{m}_i ) + 2 y_\Lambda N,
\label{eq:effActionSineGordon}
\end{align}
which (up to the field-independent terms) corresponds to the scale-dependent local potential of the sine-Gordon field theory.
Obviously, the parameter $y_{\Lambda}$ satisfies the same 
flow equation (\ref{eq:flowun}) as the couplings $u_{\Lambda}^{(2n)}$, 
\begin{equation}
  \partial_{l} y_{\Lambda}  =  
  - \frac{\pi}{    \tau_{\Lambda}}  y_{\Lambda}  +
 {\cal{O}} ( y_{\Lambda}^2 ).
 \label{eq:flowyres}
 \end{equation}
Introducing  the dimensionless rescaled coupling
 \begin{equation}
 \tilde{y}_l = \frac{ ( 2 \pi )^3 2 y_{\Lambda}}{ c_{\Lambda} \Lambda^2 }
 = - \frac{u^{(4)}_{\Lambda}}{ 2 \pi c_{\Lambda} \Lambda^2}
,
 \label{eq:tildey}
 \end{equation}
we obtain from Eq.~(\ref{eq:flowyres})
 \begin{equation}
 \partial_l \tilde{y}_l = ( 2 - \eta_l - \pi / \tau_{l} ) \tilde{y}_l + {\cal{O}} ( \tilde{y}_l^2 ).
 \label{eq:flowy2}
 \end{equation}
Here
the flowing anomalous dimension of the vortex field $\bar{m}_{\bd{k}}$  is defined by
 \begin{equation}
 \eta_l = \frac{ \partial_l \tau_\Lambda }{\tau_\Lambda }.
 \label{eq:etadef}
 \end{equation}
We shall show later that $\eta_l \propto \tilde{y}_l^2$, so that
to linear order in $\tilde{y}_l$ we may neglect the contribution of $\eta_l$
on the right-hand side of Eq.~(\ref{eq:flowy2}), which then reduces to the well-known
RG equation for the vortex fugacity of the XY-model \cite{Kosterlitz73},
\begin{equation}
 \partial_l \tilde{y}_l = ( 2 - \pi / \tau_{l} ) \tilde{y}_l + {\cal{O}} ( \tilde{y}_l^2 ).
 \label{eq:flowy3}
 \end{equation}
Note that the right-hand side of the flow equation (\ref{eq:flowy3}) vanishes for
 \begin{equation}
 \tau_{\ast} = \pi /2.
 \end{equation}
The corresponding RG fixed point describes the 
BKT phase transition at $T_c  = \tau_{\ast} J = \pi J /2$. 

\subsection{Flow equation for the temperature}
\label{sec:xyFlowMarginal}

Next, let us derive an infinite hierarchy of RG flow equations for the set of
marginal couplings $c_{\Lambda}^{(2n)}$
defined via the long-wavelength expansion (\ref{eq:vertexexp}) of the vertices
$\Gamma^{(2n)}_{\Lambda} ( \bd{k} , - \bd{k} , 0, \ldots , 0 )$, from which we then recover the
well-known RG equation for the scale-dependent dimensionless temperature
$\tau_\Lambda = c_{\Lambda} / a^2 = c_{\Lambda}^{(2)}/2$.
To derive this hierarchy, let us first consider the RG flow of
$c_{\Lambda} $ which is given by the exact FRG
flow equation (\ref{eq:cexactflow}). Since we can neglect, to leading order in $y_\Lambda$, the loop momentum $\bd{q}$ which appears in the
argument of the single-scale propagator $\dot{G}_{\Lambda} ( \bd{q} )$ on the right-hand side, the RG flow of $c_{\Lambda}$ reduces to  
 \begin{equation}
 \partial_{\Lambda} c_{\Lambda} = \frac{ a^2}{4 } A_{\Lambda} 
   c^{(4)}_{\Lambda}  ,
 \label{eq:cflow}
 \end{equation}
where $A_{\Lambda}$ is defined in Eq.~(\ref{eq:Adef})
and the marginal coupling $c_{\Lambda}^{(4)}$ can be written as
 \begin{equation}
 c^{(4)}_{\Lambda}  = a^{-2} \lim_{k \rightarrow 0} \partial_k^2 
 \Gamma^{(4)}_{\Lambda}  (     \bd{k} ,  - \bd{k} ,  0,0   ),
 \label{eq:c4def}
 \end{equation}
which is a special case of the definition~(\ref{eq:cndef}) for $n=2$.
Approximating  the factor $A_{\Lambda}$ as described in Eq.~(\ref{eq:Adef}) 
we obtain
 \begin{equation}
 \partial_{l} \tau_{\Lambda}  = \frac{ c_{\Lambda}^{(4)}}{ 8 \pi
 \tau_{\Lambda} },
 \label{eq:flowtau}
 \end{equation}
where $\partial_l = - \Lambda \partial_\Lambda$ denotes again the logarithmic scale derivative.
To determine the RG flow of the scale-dependent dimensionless temperature $\tau_{\Lambda}$,
we need the flow of the marginal coupling $c_{\Lambda}^{(4)}$ related to the
momentum dependence of the four-point vertex. 
Differentiating Eq.~(\ref{eq:flow4b}) twice with respect to $k$ and taking the limit $k \to 0$ we obtain
\begin{align}
\partial_{\Lambda} c_{\Lambda}^{(4)} &= \frac{ c_{\Lambda}^{(6)}}{2}  A_{\Lambda} - 5 u^{(4)}_\Lambda c^{(4)}_\Lambda B_\Lambda (0)
\nonumber
\\
&- 2 \left( u^{(4)}_{\Lambda} \right)^2  a^{-2}  \lim_{ k \rightarrow 0}  \partial_k^2    B_{\Lambda} ( k ),
\end{align}
where
 \begin{equation}
 B_{\Lambda} ( k ) =
 \frac{1}{N}
 \sum_{\bd{q}}
 \dot{G}_{\Lambda} ( \bd{q} ) G_{\Lambda} ( \bd{q} + \bd{k} ).
 \label{eq:definitionBLambdaK}
 \end{equation}
With our cutoff scheme, the expansion of this integral for small $k$
can be obtained analytically,
 \begin{equation}
 B_{\Lambda} ( k ) = B_{\Lambda}^0 +    \frac{ k^2 a^2}{2} B_{\Lambda}^{\prime \prime}
 + {\cal{O}} ( k^4 ),
 \end{equation}
where
 \begin{align}
 B_{\Lambda}^0 &=
 - \frac{ a^2 \left( \Lambda^3 c_{\Lambda} + \frac{1}{4} \Lambda^4 \partial_\Lambda c_\Lambda \right) }{ 2\pi ( r_{\Lambda} + c_{\Lambda} \Lambda^2 )^3 }
 \approx - \frac{a^2}{2 \pi   c_{\Lambda}^2  \Lambda^3 },
 \label{eq:BLambdaK0}
 \\
 B_{\Lambda}^{\prime \prime} &=
 \frac{ c_{\Lambda}^2 \Lambda^3}{ 2\pi ( r_{\Lambda} + c_{\Lambda} \Lambda^2 )^4 }
  \approx \frac{1}{2 \pi   c_{\Lambda}^2  \Lambda^5 },
 \end{align}
and hence
 \begin{equation}
 a^{-2} \lim_{ k \rightarrow 0}  \partial_k^2    B_{\Lambda} ( k ) = B_{\Lambda}^{\prime \prime}
 \approx
  \frac{1}{2 \pi   c_{\Lambda}^2  \Lambda^5 }.
 \end{equation}
Using $c_{\Lambda} = \tau_{\Lambda} a^2$ we find that the
dimensionless coupling $c^{(4)}_{\Lambda}$ satisfies the flow equation
 \begin{equation}
   \partial_{l} c_{\Lambda}^{(4)} =  \frac{ c_{\Lambda}^{(6)}}{ 4 \pi
 \tau_{\Lambda}} - \frac{5 u^{(4)}_\Lambda c^{(4)}_\Lambda}{2\pi \tau_\Lambda c_\Lambda \Lambda^2} + \frac{1}{\pi} \left(
 \frac{ u^{(4)}_{\Lambda}}{ c_{\Lambda} \Lambda^2 } \right)^2.
 \label{eq:c4flow}
 \end{equation}
In terms of the dimensionless coupling
$\tilde{y}_l$ defined in Eq.~(\ref{eq:tildey}),
the flow equation (\ref{eq:c4flow}) can be written as
 \begin{equation}
   \partial_{l} c_\Lambda^{(4)} =  \frac{ c_\Lambda^{(6)}}{ 4 \pi
 \tau_\Lambda} + \frac{5 c^{(4)}_\Lambda \tilde{y}_l}{\tau_\Lambda} +  4 \pi  \tilde{y}_l^2.
 \label{eq:c4flowl}
 \end{equation}
Anticipating that all marginal couplings $c^{(2n)}_\Lambda$ for $n \geq 2$ are of the order $y_\Lambda^2$, we see that to leading order in the fugacity we can neglect the second term in Eq.~\eqref{eq:c4flowl}, so that
 \begin{equation}
   \partial_{l} c_\Lambda^{(4)} =  \frac{ c_\Lambda^{(6)}}{ 4 \pi
 \tau_\Lambda} +  4 \pi  \tilde{y}_l^2.
 \label{eq:c4flowl2}
 \end{equation}
Obviously, the RG flow of $ c_{\Lambda}^{(4)}$ depends 
on the coupling $c_{\Lambda}^{(6)}$ related to the momentum-dependent part of the
six-point vertex and 
on the dimensionless coupling $\tilde{y}_l$ related to the momentum-independent part of the vertices, whose flow equation to leading order in the fugacity is given in Eq.~(\ref{eq:flowy3}).
To identify the general structure of the infinite hierarchy of flow equations for the marginal couplings $c^{(2n)}_\Lambda$,
let us also write down the flow equations for
$n=3$ and $n=4$ to leading order.
For $n=3$ we find
 \begin{align}
 \partial_{\Lambda} c_{\Lambda}^{(6)} &= \frac{ c^{(8)}_{\Lambda}}{2} 
  A_{\Lambda}
   -  8 u_{\Lambda}^{(4)} u_{\Lambda}^{(6)} B_{\Lambda}^{\prime \prime} .
 \end{align}
Inserting our previous results for $A_\Lambda$ and for $B_\Lambda''$, 
expressing all $u_\Lambda^{(2n)}$ in terms of $\tilde{y}_l$ via Eqs.~(\ref{eq:undef}) and (\ref{eq:tildey}), 
and finally using again $l = \ln ( \Lambda_0 / \Lambda )$
as flow parameter, we obtain for the flow of the marginal part of the
six-point vertex
  \begin{equation}
   \partial_{l} c_\Lambda^{(6)} =  \frac{ c_\Lambda^{(8)}}{ 4 \pi
 \tau_\Lambda} -  ( 4 \pi)^3 \tilde{y}_l^2.
 \label{eq:c6flowl}
 \end{equation}
In the next order $n=4$ we get
\begin{align}
 \partial_{\Lambda} c_{\Lambda}^{(8)} &= \frac{ c^{(10)}_{\Lambda}}{2} 
 A_{\Lambda}
  - \bigl[ 12 u_{\Lambda}^{(4)} u_{\Lambda}^{(8)} + 20 ( u_{\Lambda}^{(6)} )^2 \bigr]
 B_{\Lambda}^{\prime \prime},
 \end{align}
which results in
 \begin{equation}
   \partial_l c_\Lambda^{(8)} =  \frac{ c_\Lambda^{(10)}}{ 4 \pi
 \tau_\Lambda} +  ( 4 \pi)^5  \tilde{y}_l^2.
 \label{eq:c8flowl}
 \end{equation}
By comparing the RG equations for $c_\Lambda^{(4)}$, $c_\Lambda^{(6)}$, and $c_\Lambda^{(8)}$
given in
Eqs.~(\ref{eq:c4flowl2}), (\ref{eq:c6flowl}), and (\ref{eq:c8flowl}) 
we conclude that at least for  $n=2,3,4$ the flow of the marginal couplings $c_l^{(2n)}$
is given by
 \begin{equation}
   \partial_{l} c_{l}^{(2n)} =  \frac{ c_{l}^{(2n+2)}}{ 4 \pi
 \tau_{l}} + (-1)^n ( 4 \pi)^{2n-3}  \tilde{y}_l^2,
 \label{eq:cnflowl}
 \end{equation}
where, with a slight abuse of notation, from now on all couplings
are considered as functions of the
logarithmic flow parameter $l$.
By carefully examining the combinatorics
in the exact FRG flow equations for the marginal couplings $c^{(2n)}_l$ for arbitrary $n$, we show in Appendix~\ref{sec:appFlowMarginalAllOrders} that  Eq.~(\ref{eq:cnflowl}) describes the flow of the marginal couplings $c^{(2n)}_l$ to second order in $y_\Lambda$ for arbitrary integer $n \geq 2$. 

\begin{figure}
\includegraphics[width=\columnwidth]{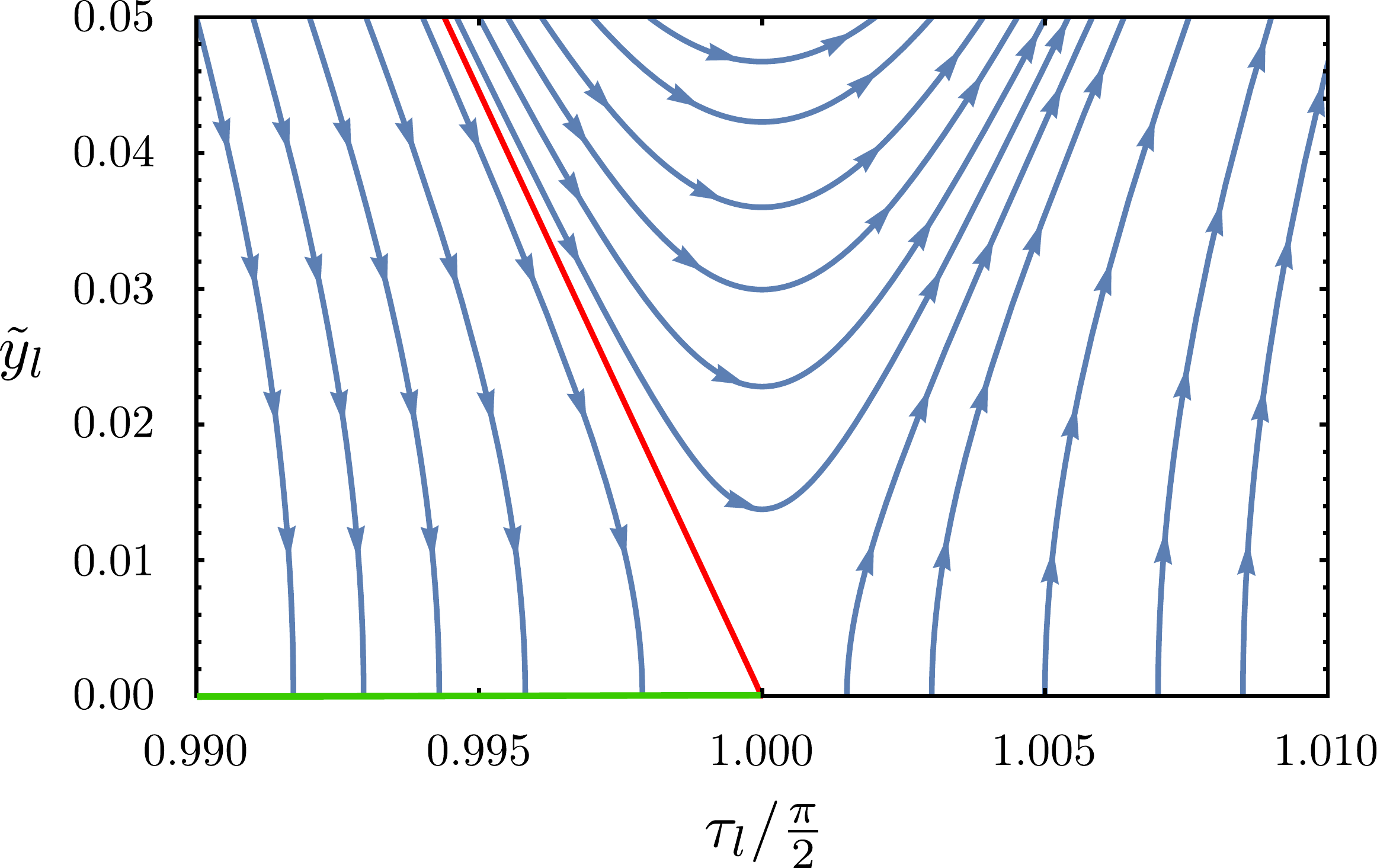}
\caption{
%(Color online)
Flow diagram for the XY-model in the vicinity of the BKT transition at $\tau_\ast = T_c / J = \pi / 2$, which was obtained by numerically solving the flow equations (\ref{eq:flowtaures}) and (\ref{eq:flowy4}). The separatrix (red line) separates the high-temperature regime on the right from the low-temperature regime on the left. In the disordered high-temperature regime $\tilde{y}_l$ and $\tau_l$ flow to infinity, while in the quasi-long-range ordered low-temperature regime the flow terminates at the Gaussian fixed point manifold (green line) at $\tilde{y} = 0$ and $\tau \leq \pi / 2$.
}
\label{fig:xy_schematic_flow}
\end{figure}

The infinite system of RG equations for the marginal couplings given above can be solved by the recursion
 \begin{equation}
 c_l^{(2n+2)} =  - ( 4 \pi)^2 c_l^{(2n)}, \quad n \geq 2.
 \label{eq:crec}
 \end{equation}
Substituting this into Eq.~(\ref{eq:cnflowl}), we obtain a set of decoupled RG equations
for the marginal couplings $c_l^{(2n)}$,
 \begin{equation}
   \partial_{l} c_{l}^{(2n)} =  - \frac{ 4 \pi }{ \tau_l} c_{l}^{(2n)}
   + (-1)^n ( 4 \pi)^{2n-3}  \tilde{y}_l^2.
 \label{eq:cndecoupled}
 \end{equation}
The solution to these equations with initial condition $c_{l=0}^{(2n)} =0$ is
 \begin{align}
 &c^{(2n)}_l = (-1)^n (4 \pi)^{2n-3} \int_0^l d l^{\prime} \tilde{y}_{l^{\prime}}^2
 e^{- 4 \pi \int_{l^{\prime}}^l dx / \tau_{x} }
 \nonumber
 \\
 &= (-1)^n (4 \pi)^{2n-3} \tilde{y}_0^2 \int_0^l d l^{\prime} e^{\int_0^{l'} dx (4 - 2\pi/\tau_x) - 4 \pi \int_{l^{\prime}}^l dx / \tau_{x} },
 \label{eq:c2nint}
 \end{align}
where in the second line we have used the formal solution to Eq.~\eqref{eq:flowy3}.
One easily verifies that this expression for $c^{(2n)}_l$ indeed satisfies
the recursion relation (\ref{eq:crec}).
From Eq.~\eqref{eq:c2nint} we see that $c_l^{(4)}$ and thus also $\partial_l \tau_l$ are proportional to $\tilde{y}_0^2$, so that to leading order we may replace the scale-dependent dimensionless temperature by its initial value $\tau$. The integration is then elementary
and we obtain
 \begin{equation}
 c^{(2n)}_l = (-1)^n \frac{(4 \pi)^{2n-3} \tilde{y}^2_l}{4 + 2\pi / \tau} \left[ 1 - e^{-(4 + 2\pi / \tau) l} \right],
 \end{equation}
so that the marginal couplings
$c^{(2n)}_l$ with $n \geq 2$ are indeed proportional to $\tilde{y}_l^2$ as anticipated.
For large $l \gtrsim (4 + 2\pi / \tau)^{-1}$ we can neglect the $e^{-(4 + 2\pi / \tau) l}$ contribution,
 \begin{equation}
 c^{(2n)}_l = (-1)^n \frac{(4 \pi)^{2n-3} \tilde{y}^2_l}{4 + 2\pi / \tau},
 \label{eq:resultMarginalGeneral}
 \end{equation}
so that for $n=2$ we find
\begin{equation}
 c^{(4)}_l = \frac{4\pi}{4 + 2\pi / \tau} \tilde{y}^2_l.
 \label{eq:c4res}
 \end{equation}
Substituting this into the flow equation (\ref{eq:flowtau}) we obtain for the
flow of the dimensionless temperature
 \begin{equation}
 \partial_{l} \tau_{l} = \frac{\tilde{y}_l^2}{8 \tau + 4\pi} + {\cal{O}} ( \tilde{y}_l^3 ),
 \end{equation}
which close to the BKT transition at $\tau_{\ast} = \pi / 2$ simplifies to
\begin{equation}
 \partial_{l} \tau_{l} = \frac{\tilde{y}_l^2}{8\pi} + {\cal{O}} ( \tilde{y}_l^3 ).
 \label{eq:flowtaures}
 \end{equation}
This equation should be solved simultaneously with the
 flow equation for $\tilde{y}_l$, which according to Eq.~(\ref{eq:flowy3}) is given by
 \begin{equation}
 \partial_l \tilde{y}_l = ( 2  - \pi / \tau_{l} ) \tilde{y}_l + {\cal{O}} ( \tilde{y}_l^2 ).
 \label{eq:flowy4}
 \end{equation}
Eqs.~(\ref{eq:flowtaures}) and (\ref{eq:flowy4})
are the well-known Kosterlitz-Thouless RG equations for the two-dimensional
XY-model \cite{Kosterlitz73,Chaikin95,Herbut07}. The corresponding flow diagram is shown in Fig. \ref{fig:xy_schematic_flow}.
Note that the numerical coefficient in front of $\tilde{y}_l^2$
in the flow equation (\ref{eq:flowtaures})
depends on the specific cutoff procedure and
can be arbitrarily changed by rescaling the coupling $\tilde{y}_l$ by a finite numerical factor
\footnote{To obtain the same prefactor as in the original work by Kosterlitz and Thouless \cite{Kosterlitz73} (which is also used in the textbook by Chaikin and Lubensky \cite{Chaikin95}) we should redefine {$\tilde{y}_l \rightarrow \sqrt{2} ( 2 \pi )^2 \tilde{y}_l$}, in which case Eq.~(\ref{eq:flowy4}) takes the form $ \partial_{l} \tau_{l} = 4 \pi^3 \tilde{y}_l^2.$ On the other hand, the coupling {$\hat{y}_l$} introduced by Herbut \cite{Herbut07} corresponds in our notation to {$\tilde{y}_i = 2 \sqrt{2} \hat{y} $}; the corresponding flow equation is $\partial_{l} \tau_{l} = \hat{y}_l^2 / \pi$.}. Furthermore we observe that, while Eq.~\eqref{eq:flowy4} is of first order in $\tilde{y}_l$, it really is of second order in the double expansion in $(\tau - \tau_\ast)$ and $\tilde{y}_l$, so that we should also evaluate the $\tilde{y}_l^2$ correction at $\tau = \tau_\ast$. However, an explicit calculation shows (see Appendix~\ref{sec:appTechDetails} for technical details) that this correction vanishes, so that our result \eqref{eq:flowy4} is indeed correct to second order in the double expansion in $(\tau - \tau_\ast)$ and $\tilde{y}_l$.

\section{Effect of amplitude fluctuations on the 
BKT transition}
\label{sec:amplitudeFlucts}

Having derived the Kosterlitz-Thouless flow equations for the XY-model within the FRG, 
we are now in a position to assess the effect of amplitude fluctuations on the BKT transition. 
As already mentioned  in Sec.~\ref{sec:intro}, 
the question of how amplitude fluctuations influence the BKT transition
is not fully settled since 
previous FRG calculations for the $O(2)$-model have so far not been able to 
encounter the  expected line of fixed points without fine-tuning the regulator. 
In this  section, we will address this question by considering a bosonic lattice model 
close to the supposed phase transition, which in the limit of a hard-core contact 
interaction reduces to the XY-model. 
Assuming weak amplitude fluctuations, we show that their effect can be described via an effective vortex 
interaction. Evaluating the effect of this effective interaction on our RG flow equations for the 
XY-model enables us to conclude that amplitude fluctuations do not destroy the BKT phase transition.

\subsection{Adding amplitude fluctuations to the XY-model}

\subsubsection{Bosonic lattice action in Villain approximation}

To clarify the influence of amplitude fluctuations on the BKT transition, let us start from 
a lattice model  describing  bosons in the grand canonical ensemble at temperature $T$, 
which interact via a repulsive contact interaction $U > 0$,
\begin{align}
{S} [\psi] = \frac{U}{2T} \sum_i \left( \left| \psi_i \right|^2 - \alpha^2 \right)^2 + \frac{J}{2 T} \sum_{i \mu} \left| \psi_{i+\mu} - \psi_i \right|^2,
\label{eq:amplitudeFullAction}
\end{align}
where $\psi_i$ describes the complex bosonic field at site $i$ of a square lattice with lattice spacing $a$, the exchange interaction is denoted by $J > 0$, and we have neglected the non-zero Matsubara frequencies as we are interested in the regime close to the classical (finite temperature) BKT transition. We assume that $\alpha > 0$ and that $U$ is sufficiently large, so that the amplitude fluctuations of the bosonic field are small compared with the radius $\alpha$ of the Mexican hat potential given by the first term in Eq.~\eqref{eq:amplitudeFullAction}. Writing the complex field in density-phase notation,
\begin{align}
\psi_i = \sqrt{\rho_i}e^{i \theta_i},
\end{align}
we specifically assume that $U \bar{\rho}^2 / T \gg 1$, where $\bar{\rho} = \left< \rho_i \right>$ is the expectation value of the $\rho$-field. Rewriting the action as a functional of density and phase we find
\begin{align}
&{S} [\rho, \theta] = \frac{U}{2T} \sum_i \left( \rho_i - \alpha^2 \right)^2
\nonumber
\\
&+ \frac{J}{2 T} \sum_{i \mu} \left[ \rho_i + \rho_{i+\mu} - 2 \sqrt{\rho_i \rho_{i+\mu}} \cos (\theta_{i+\mu} - \theta_i) \right].
\label{eq:ampDensPhaseExact}
\end{align}
Note that in the limit $U \to \infty$ we can replace the field $\rho_i$ in the action by its expectation value $\bar{\rho}$. Redefining $J \bar{\rho} \to J$ and neglecting constant terms we then recover the action \eqref{eq:actionXYmodel} of the XY-model. A large but finite $U$ thus allows us to study the effect of amplitude fluctuations on the BKT transition. Analogous to the Villain approximation for the XY-model we  expand the exponential of the cosine as
\begin{align}
&\indent \exp \left[ \frac{J \sqrt{\rho_i \rho_{i+\mu}}}{T} \cos (\Delta_\mu \theta_i) \right]
\nonumber
\\
&\approx R_{V,i \mu} \sum_{n_{i \mu} = -\infty}^\infty \exp \left[ - \frac{
 (\Delta_\mu \theta_i - 2\pi n_{i \mu})^2  }{2 \tau_{V,i\mu}}      \right],
 \label{eq:Villain2}
\end{align}
where we have used $\Delta_\mu \theta_i = \theta_{i+\mu} - \theta_i$ [see Eq.~\eqref{eq:defThetaDiff}]
and defined
\begin{subequations}
\begin{align}
R_{V,i\mu} = R_V (\sqrt{\rho_i \rho_{i+\mu}} / \tau ),
 \label{eq:RVidef}
\\
\tau_{V,i\mu} = \tau_V ( \sqrt{\rho_i \rho_{i+\mu}} / \tau).
\end{align}
\end{subequations}
Here $\tau = J/T$ and the functions $R_V$ and $\tau_V$ are defined in Eqs.~\eqref{eq:VillainR} and \eqref{eq:VillainTau}. We may now separate the contribution from the phase fluctuations to the partition function as follows,
\begin{align}
Z_\text{amp} &= \prod_i \left( \int_0^\infty d\rho_i \right) \exp \Bigg[ - \frac{U}{2T} \sum_i \left( \rho_i - \alpha^2 \right)^2 
\nonumber
\\
&  \hspace{10mm} - \frac{2J}{T} \sum_i \rho_i    + \sum_{i \mu} \ln R_{V,i\mu} \Bigg] Z_\text{eff} [\rho],
\end{align}
where
\begin{align}
Z_\text{eff} [\rho] &= \prod_i \left( \int_0^{2\pi} \frac{d\theta_i}{2\pi} \right) \prod_{i\mu} \left( \sum_{n_{i\bd{\mu}} = -\infty}^\infty \right)
\nonumber
\\
&\times \exp \left[ - \sum_{i \mu} \frac{(\Delta_\mu \theta_i - 2\pi n_{i\mu})^2}{2 \tau_{V,i\mu}} \right].
\label{eq:zThetaVillainAmp}
\end{align}
Although $\tau_{V,i\mu}$ depends on $i$ and $\mu$, we find that we can still apply the same transformations we have used in the XY-model between Eqs.~\eqref{eq:ZVillain} and \eqref{eq:ZVillain3} to arrive at
\begin{align}
Z_\text{eff} [\rho] &= \exp \left[ \frac{1}{2} \sum_{i \mu} \ln \tau_{V,i\mu} \right] \prod_{i \mu} \left( \sum_{p_{i\mu} = - \infty}^\infty \right) \delta_{\bd{\Delta} \cdot \bd{p}_i,0}
\nonumber
\\
& \hspace{3mm} \times \exp \left[ - \frac{1}{2} \sum_{i \mu} \tau_{V,i\mu} p_{i \mu}^2 \right],
\label{eq:ampFlucCurrentForm}
\end{align}
where the lattice divergence $\bd{\Delta} \cdot \bd{p}_i$ is defined in Eq.~\eqref{eq:defPDelta}.

\subsubsection{Expansion in density fluctuations}

Since we have assumed that the density fluctuations are small compared to the expectation value $\left< \rho_i \right> = \bar{\rho}$ of the density field, it is reasonable to define
\begin{align}
\rho_i = \bar{\rho} ( 1 + \rt_i )
\end{align}
and to expand our action in terms of the density fluctuations $\rt_i$. For $\tau_{V,i\mu}$ this yields
\begin{align}
\tau_{V,i\mu} &= \tau_V^{(0)} + \tau_V^{(1)} \frac{\rt_{i+\mu} + \rt_i}{2} + \frac{\tau_V^{(2)}}{2} \left( \frac{\rt_{i+\mu} + \rt_i}{2} \right)^2
\nonumber
\\
&
- \frac{\tau_V^{(2-)}}{2} \left( \frac{\rt_{i+\mu} - \rt_i}{2} \right)^2 + \mathcal{O} (\rt^3),
\label{eq:tauExpansionAmp}
\end{align}
where the coefficients are given by (we use again the notation $\tau = T / J $)
\begin{subequations}
\begin{align}
\tau_V^{(0)} &= \tau_V \left(  \frac{\bar{\rho}}{  \tau } \right),
\\
\tau_V^{(1)} = \tau_V^{(2-)} &= \frac{ \bar{\rho}}{\tau} \tau_V' \left( \frac{ \bar{\rho}}{\tau} \right),
\\
\tau_V^{(2)} &= \left( \frac{ \bar{\rho}}{\tau} \right)^2 \tau_V'' \left( \frac{ \bar{\rho}}{\tau} \right).
\end{align}
\end{subequations}
The reason for also taking the second-order terms in $\rt$ into account is that this allows us to extend the integrations over the density fluctuations to the entire real axis. Following the transformation outlined in Sec.~\ref{sec:dualRepresentations} [see Eqs.~\eqref{eq:defMfieldA} - \eqref{eq:defMfieldD}], we now introduce another set of integers $m_i$ to replace the $p_{i \mu}$-field. With the notation
\begin{align}
M_i = \frac{1}{2} \sum_{{\mu}} \left[ \left( m_{i+x+y} - m_{i+x+y-\mu} \right)^2 + \left( m_{i+\mu} - m_i \right)^2 \right]
\label{eq:defM}
\end{align}
we may then write
\begin{align}
&Z_\text{eff} [\rho] = \exp \left[ \frac{1}{2} \sum_{i \mu} \ln \tau_{V,i\mu} \right] \prod_i \left( \sum_{m_i = - \infty}^\infty \right)
\nonumber
\\
&\times \exp \left\{ - \frac{1}{2} \sum_i  M_i
 \left[ \tau_V^{(0)}  + \tau_V^{(1)}   \rt_i
+  \frac{\tau_V^{(2)}}{2} \rt_i^2  \right] \right\},
\end{align}
where we have approximated $\rt_{i+\mu} \approx \rt_i$ in the $\rt^2$ term, since it is only needed to ensure convergence and close to the BKT transition it is sufficient to consider the long-wavelength limit.
With the expansion (\ref{eq:tauExpansionAmp})
of $\tau_{V,i\mu}$ we can also expand
\begin{align}
\frac{1}{2} \sum_{i\mu} \ln \tau_{V,i\mu} = \kappa_V^{(0)} + \kappa_V^{(1)} \sum_i \rt_i 
+ \frac{ \kappa_V^{(2)}}{2} \sum_i \rt_i^2 + \mathcal{O} (\rt^3),
\end{align}
where $\kappa_V^{(0)}$ is an uninteresting constant and
\begin{subequations}
\begin{align}
\kappa_V^{(1)} &= \frac{\tau_V^{(1)}}{\tau_V^{(0)}},
\\
\kappa_V^{(2)} &= \frac{\tau_V^{(2)}}{\tau_V^{(0)}} - \left( \frac{\tau_V^{(1)}}{\tau_V^{(0)}} \right)^2,
\end{align}
\end{subequations}
and we have again approximated $\rt_{i+\mu} \approx \rt_i$ in the second-order term. 
Analogously, we expand the coefficient
$R_{V,i\mu}$ defined in Eq.~(\ref{eq:RVidef}),
\begin{align}
R_{V,i\mu} = R_V^{(0)} + R_V^{(1)} \frac{\rt_{i+\mu} + \rt_i}{2} + \frac{R_V^{(2)}}{2} \rt_i^2 + \mathcal{O} (\rt^3),
\end{align}
where
\begin{subequations}
\begin{align}
R_V^{(0)} &= R_V \left( \frac{\bar{\rho}}{\tau} \right),
\\
R_V^{(1)} &= \left( \frac{ \bar{\rho}}{\tau} \right) R_V' \left( \frac{ \bar{\rho}}{\tau} \right),
\\
R_V^{(2)} &= \left( \frac{ \bar{\rho}}{\tau} \right)^2 R_V'' \left( \frac{ \bar{\rho}}{\tau} \right),
\end{align}
\end{subequations}
so that we obtain for the logarithm of $R_{V,i\mu}$
\begin{align}
\sum_{i\mu} \ln R_{V,i\mu} = L_V^{(0)} + 2 L_V^{(1)} \sum_i \rt_i + L_V^{(2)} \sum_i \rt_i^2 + \mathcal{O} (\rt^3),
\end{align}
where $L_V^{(0)}$ is again an unimportant constant and
\begin{subequations}
\begin{align}
L_V^{(1)} &= \frac{R_V^{(1)}}{R_V^{(0)}},
\\
L_V^{(2)} &= \frac{R_V^{(2)}}{R_V^{(0)}} - \left( \frac{R_V^{(1)}}{R_V^{(0)}} \right)^2.
\end{align}
\end{subequations}
Putting everything together we find that the full partition function can be written as
\begin{align}
Z_\text{amp} &= \prod_i \left( \int_{-\infty}^\infty d\rt_i \sum_{m_i = -\infty}^\infty \right) \exp \Bigg\{ -  \sum_i  \bigg[ \frac{u}{2} \rt_i^2 + v  \rt_i
\nonumber
\\
& \hspace{10mm} + \frac{\tau_V^{(0)}}{2}  M_i + \frac{\tau_V^{(1)}}{2}  \rt_i M_i + \frac{\tau_V^{(2)}}{4}  \rt_i^2 M_i  \bigg] \Bigg\},
\label{eq:partitionFuncDensApprox}
\end{align}
where we have defined
\begin{align}
u  &= \frac{U \bar{\rho}^2}{T} - 2 L_V^{(2)} - \kappa_V^{(2)} \approx \frac{U \bar{\rho}^2}{T},
\\
v &= \frac{2 \bar{\rho}}{\tau} + \frac{U \bar{\rho} (\bar{\rho} - \alpha^2)}{T} - 2 L_V^{(1)} - 
\kappa_V^{(1)}.
\end{align}
Let us briefly consider the order of magnitude of the couplings in our new action \eqref{eq:partitionFuncDensApprox}. Close to the BKT transition, $\bar{\rho} / \tau$ is of order unity so that $\tau_V^{(0)}$, $\tau_V^{(1)}$, and $\tau_V^{(2)}$ are also of order unity, while $u \gg 1$ due to the dominance of the first term in the definition of $u$. Using the fact that $\left< \rt_i \right>$ vanishes by construction, it is easy to show that
\begin{align}
v + \frac{\tau_V^{(1)}}{2} \left< M_i \right> = 0.
\end{align}
Since $\left< M_i \right>$ is of order unity below the BKT transition, it follows that $v$ is also of order unity.

\subsubsection{Integrating out the density fluctuations}

In order to make contact with our calculations for the XY-model, we now integrate out the $\rt$-field to derive an effective theory for the dual vortex field $m$. Performing the Gaussian integrals and neglecting constant terms in the action yields
\begin{align}
Z_\text{amp} &= \prod_i \left( \sum_{m_i = -\infty}^\infty \right) \exp \Bigg\{ - \frac{1}{2} \sum_i 
 \biggl[ \tau_V^{(0)}   M_i
\nonumber
\\
& +  \ln \Bigl( 1 + \frac{\tau_V^{(2)}}{2 u} M_i \Bigr) -  \frac{\Bigl( v + \frac{\tau_V^{(1)}}{2} M_i \Bigr)^2}{u + \frac{\tau_V^{(2)}}{2 } M_i}\biggr] \Bigg\}.
\label{eq:ampIntegratedOut}
\end{align}
For large $M_i$ the argument of the exponential in Eq.~\eqref{eq:ampIntegratedOut} can be approximated as
\begin{align}
-{S} [m] \approx - \frac{1}{2} \biggl[ \tau_V^{(0)} - \frac{\bigl( \tau_V^{(1)} \bigr)^2}{2 \tau_V^{(2)}} 
\biggr] \sum_i M_i.
\end{align}
Since the term in the square brackets is always positive and, close to the BKT transition, 
of order unity, we find that the partition function \eqref{eq:ampIntegratedOut} 
is well defined and that the internal sums over $m_i$ are effectively 
cut off at $M_i \approx 1$, so that we may approximate
\begin{align}
Z_\text{amp} &= \prod_i \left( \sum_{m_i = -\infty}^\infty \right) \exp \Bigg[ - \frac{\tau^{\prime}}{2} \sum_i M_i
\nonumber
\\
& \hspace{20mm}
 + \frac{(\tau_V^{(1)})^2}{8 } \sum_i 
\frac{M_i^2}{u + \frac{\tau_V^{(2)}}{2} M_i} 
\Bigg],
\end{align}
where we have introduced the effective dimensionless temperature
\begin{align}
\tau^{\prime} = \tau_V^{(0)} + \frac{\tau_V^{(2)}}{2 u} - \frac{v \tau_V^{(1)}}{u} \approx \tau_V^{(0)}.
\end{align}
Expanding for large  $u$,
\begin{align}
\exp \left[ \frac{(\tau_V^{(1)})^2}{8} \sum_i 
\frac{ M_i^2}{u + \frac{\tau_V^{(2)}}{2} M_i} 
 \right] \approx 1 + \frac{(\tau_V^{(1)} )^2}{8u}
 \sum_i M_i^2,
\end{align}
and defining the dimensionless coupling constant
\begin{align}
g = \frac{3 \bigl( \tau_V^{(1)} \bigr)^2}{u},
\end{align}
we finally arrive at
\begin{align}
Z_\text{amp} &= \prod_i \left( \sum_{m_i = -\infty}^\infty \right) \exp \left[ - \frac{\tau^{\prime}}{2} \sum_i M_i \right]
\nonumber
\\
&\hspace{20mm} \times \left( 1 + \frac{g}{4!} \sum_i M_i^2 \right).
\label{eq:ampZ}
\end{align}
Fourier transforming to momentum space we can write this as
\begin{align}
Z_\text{amp} &= \prod_i \left( \sum_{m_i = -\infty}^\infty \right) \exp \left[ - \frac{1}{2} \sum_{\bd{k}} \omega_{\bd{k}} m_{\bd{k}} m_{-\bd{k}} \right]
\nonumber
\\
&\times \Biggl[ 1 + \frac{g}{4! N} \sum_{\bd{k}_1 \bd{k}_2 \bd{k}_3 \bd{k}_4} 
 \delta_{\bd{k}_1 + \bd{k}_2 + \bd{k}_3 + \bd{k}_4, 0}
V_{\bd{k}_1, \bd{k}_2, \bd{k}_3, \bd{k}_4}
 \nonumber
 \\
 & \hspace{30mm} \times  m_{\bd{k}_1} 
m_{\bd{k}_2} m_{\bd{k}_3} m_{\bd{k}_4} \Biggr],
\end{align}
where $\omega_{\bd{k}} = 4 \tau^{\prime} ( 1 - \gamma_{\bd{k}} ) $ is defined analogously to 
Eq.~\eqref{eq:omegakdef},  and the fully symmetrized momentum dependence of the 
effective quartic interaction due to amplitude fluctuations is given by
\begin{align}
V_{\bd{k}_1, \bd{k}_2, \bd{k}_3, \bd{k}_4} = \frac{1}{3} 
 \left[ \tilde{V}_{\bd{k}_1, \bd{k}_2; \bd{k}_3, \bd{k}_4} + 
 ( \bd{k}_2 \leftrightarrow  \bd{k}_3 ) +  ( \bd{k}_2 \leftrightarrow \bd{k}_4 ) \right],
\label{eq:VSymmetrized}
\end{align}
where
\begin{align}
\tilde{V}_{\bd{k}_1, \bd{k}_2; \bd{k}_3, \bd{k}_4} &= \frac{1}{4} \sum_{\mu \nu \in \{x,y\}}
 \left( e^{i \bd{k}_1 \cdot \bd{a}_\mu} - 1 \right) \left( e^{i \bd{k}_2 \cdot \bd{a}_\mu} - 1 \right)
\nonumber
\\
&\indent \times   \left( 1 + e^{i (\bd{k}_1 + \bd{k}_2) \cdot (\bd{a}_x + \bd{a}_y - \bd{a}_\mu)} \right)
\nonumber
\\
&\indent \times  \left( e^{i \bd{k}_3 \cdot \bd{a}_{\nu}} - 1 \right) \left( e^{i \bd{k}_4 \cdot \bd{a}_{\nu}} - 1 \right)
\nonumber
\\
&\indent \times    \left( 1 + e^{i (\bd{k}_3 + \bd{k}_4) \cdot (\bd{a}_x + \bd{a}_y - \bd{a}_{\nu})} \right).
 \label{eq:V1234}
\end{align}

\subsection{Effect of the amplitude fluctuations on the flow equations}
\label{subsec:ampFlowEqs}

\begin{figure}
\includegraphics[width=\columnwidth]{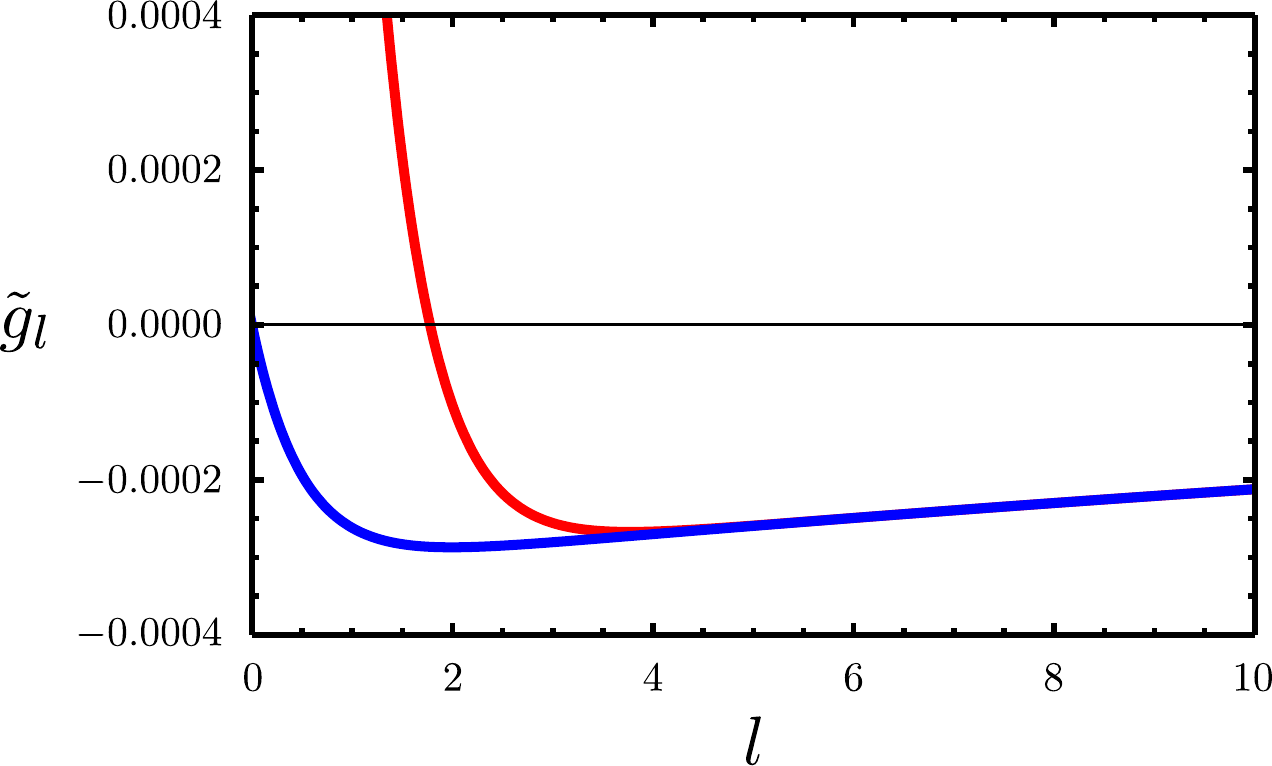}
\caption{
%(Color online)
  Representative plot of the RG flow of the irrelevant coupling $\tilde{g}_l$ 
 generated by amplitude fluctuations as a function of the logarithmic scale parameter $l = \ln ( \Lambda_0 / \Lambda )$.
The curves are obtained from the numerical solution of
the RG equations 
\eqref{eq:flowtaures}, \eqref{eq:flowy4}, and \eqref{eq:gFlowEquation}, 
with initial conditions  $\tilde{y}_0 = 0.01$ and $\tau = 0.99 \tau_\ast = 0.99 \pi / 2$. 
The upper curve (red) corresponds to a finite initial value of $\tilde{g}_0 = 0.01$ 
(corresponding to a model with finite amplitude fluctuations), 
while the lower curve (blue) corresponds to $\tilde{g}_0 = 0$ as is the case for the XY-model. 
We find that for sufficiently large $l$ both curves merge and eventually approach zero for 
$l \rightarrow \infty$, so that the large $l$ behavior 
of $\tilde{g}_l$ is independent of the initial strength of the amplitude fluctuations.
}
\label{fig:gFlow}
\end{figure}

In order to set up the FRG, we now introduce a regulator $R_\Lambda (\bd{k})$ 
analogously to the procedure described in Sec.~\ref{subsec:flow}. We thus replace
\begin{align}
\frac{1}{2} \sum_{\bd{k}} \omega_{\bd{k}} m_{\bd{k}} m_{-\bd{k}} \to \frac{1}{2} \sum_{\bd{k}} \left[ \omega_{\bd{k}} + R_\lambda (\bd{k}) \right] m_{\bd{k}} m_{-\bd{k}},
\end{align}
where
\begin{align}
R_\lambda (\bd{k}) = \zeta_\lambda \left( \lambda - \omega_{\bd{k}} \right) \Theta \left( \lambda - \omega_{\bd{k}} \right)
 \label{eq:reg2}
\end{align}
is again defined such that for $\lambda = \lambda_0$ the dispersion is constant. 
The cutoff-dependent prefactor $\zeta_\lambda$ will be specified later on; here we only demand $\zeta_{\lambda_0} = 1$. Then the cutoff-dependent generating functional $W_{\lambda_0} [h]$ 
of the connected correlation functions [see Eq.~\eqref{eq:Wdef}] at the initial scale $\lambda_0$ is given by
\begin{align}
e^{W_{\lambda_0} [h]} &=\Bigg[ 1 + \frac{g}{4! N} \sum_{\bd{k}_1 \bd{k}_2 \bd{k}_3 \bd{k}_4} 
\delta_{\bd{k}_1 + \bd{k}_2 + \bd{k}_3 + \bd{k}_4, 0} V_{\bd{k}_1, \bd{k}_2, \bd{k}_3, \bd{k}_4}
\nonumber
\\
&\hspace{12mm} \times \frac{\delta}{\delta h_{-\bd{k}_1}} \frac{\delta}{\delta h_{-\bd{k}_2}} \frac{\delta}{\delta h_{-\bd{k}_3}} \frac{\delta}{\delta h_{-\bd{k}_4}} \Bigg]
\nonumber
\\
&\indent\times
\prod_i \left( \sum_{m_i = -\infty}^\infty \right) \exp \left[ - \frac{\lambda_0}{2} \sum_i m_i^2 + \sum_i h_i m_i \right],
\end{align}
where we have pulled the quartic interaction term out of the $m$ sums by replacing the $m_{\bd{k}}$-fields 
with derivatives with respect to the source field $h_{-\bd{k}}$. This allows us to perform the sums over the $m_i$-field and expand the result to leading order in the fugacity $y_0$, see Eqs.~\eqref{eq:wim} and \eqref{eq:thetaApproxFirst}. Evaluating the derivatives with respect to $h$ is then trivial,
\begin{align}
W_{\lambda_0} [h] &= \sum_i \frac{h_i^2}{2 \lambda_0} + 2 y_0 \sum_i \cos \left( \frac{2\pi h_i}{\lambda_0} \right)
\nonumber
\\
&+ \frac{g}{4 \lambda_0^3 N} \sum_{\bd{k}} h_{\bd{k}} h_{-\bd{k}} \sum_{\bd{q}} V_{\bd{k},-\bd{k},\bd{q},-\bd{q}}
\nonumber
 \\
&+ \frac{g}{4! \lambda_0^4  N} \sum_{\bd{k}_1 \bd{k}_2 \bd{k}_3 \bd{k}_4} 
 \delta_{\bd{k}_1 + \bd{k}_2 + \bd{k}_3 + \bd{k}_4, 0}
V_{\bd{k}_1, \bd{k}_2, \bd{k}_3, \bd{k}_4} 
 \nonumber
\\
& \hspace{25mm} \times
h_{\bd{k}_1} h_{\bd{k}_2} h_{\bd{k}_3} h_{\bd{k}_4},
\end{align}
where we have expanded the right-hand side to first order in $g$ and in $y_0$. The expectation value of the $m$-field at the initial scale is therefore to leading order given by
\begin{align}
\mb_i = \frac{\delta W_{\lambda_0} [h]}{\delta h_i} = \frac{h_i}{\lambda_0} + \mathcal{O} (g, y_0),
\end{align}
so that
\begin{align}
h_i = \lambda_0 \mb_i + \mathcal{O} (g, y_0).
\end{align}
The initial condition for the average effective action $\Gamma_\lambda [\mb]$ then reads
\begin{align}
\Gamma_{\lambda_0} [\mb] &= \frac{  \tau^{\prime}  -g/\lambda_0   }{2} \sum_{\bd{k}} a^2 k^2 \left| \mb_{\bd{k}} \right|^2 - 2 y_0 \sum_i \cos (2\pi \mb_i)
\nonumber
\\
&- \frac{g}{4! N} \sum_{\bd{k}_1 \bd{k}_2 \bd{k}_3 \bd{k}_4} 
 \delta_{\bd{k}_1 + \bd{k}_2 + \bd{k}_3 + \bd{k}_4, 0}
V_{\bd{k}_1, \bd{k}_2, \bd{k}_3, \bd{k}_4} 
 \nonumber
 \\
 &\hspace{20mm} \times
\mb_{\bd{k}_1} 
 \mb_{\bd{k}_2} \mb_{\bd{k}_3} \mb_{\bd{k}_4},
\label{eq:averageEffActionAmp}
\end{align}
where we have expanded $\omega_{\bd{k}}$ to leading order and have used
\begin{align}
\frac{g}{2 \lambda_0 N} \sum_{\bd{q}} V_{\bd{k},-\bd{k},\bd{q},-\bd{q}} = \frac{g a^2 k^2}{\lambda_0} + \mathcal{O} (k^4).
\end{align}
Thus one of the effects of the amplitude fluctuations is  a correction 
to the initial value of the coupling $c_\Lambda$,
\begin{align}
c_0 = \tau^{\prime} a^2 - \frac{g a^2}{\lambda_0}.
\end{align}
Analogous to our treatment of the XY-model we now set
$\lambda = \tau^{\prime} a^2 \Lambda^2$. A convenient choice of the 
prefactor $\zeta_{\Lambda}$
in our  regulator (\ref{eq:reg2}) is then
\begin{align}
\zeta_\Lambda = \frac{c_\Lambda + \frac{g a^2}{\lambda_0}}{c_0 + \frac{g a^2}{\lambda_0}},
\end{align}
so that
\begin{align}
R_\Lambda (\bd{k}) = \left( c_\Lambda + \frac{g a^2}{\lambda_0} \right) \left( \Lambda^2 - k^2 \right) \Theta \left( \Lambda^2 - k^2 \right).
\end{align}
Although this $g$-dependent correction to the regulator 
propagates to the functions $A_\Lambda$ and $B_\Lambda (k)$
defined in Eqs.~\eqref{eq:Adef} and \eqref{eq:definitionBLambdaK}, 
it is consistent to neglect the resulting correction terms since in the derivation of the Kosterlitz-Thouless  flow equations in Sec.~\ref{sec:BKTinXYfromFRG} 
we have also dropped all corrections in $y_\Lambda$.
More important is the momentum-dependent four-point vertex in
Eq.~\eqref{eq:averageEffActionAmp}, which is explicitly given by
Eqs.~(\ref{eq:VSymmetrized}) and (\ref{eq:V1234}).
In the long-wavelength limit we obtain from Eq.~(\ref{eq:V1234})
\begin{align}
&\tilde{V}_{\bd{k}_1, \bd{k}_2; \bd{k}_3, \bd{k}_4} \approx 
\sum_{\mu \nu \in \{x,y\}} (\bd{k}_1 \cdot \bd{a}_\mu) (\bd{k}_2 \cdot \bd{a}_\mu) (\bd{k}_3 \cdot \bd{a}_\nu) (\bd{k}_4 \cdot \bd{a}_\nu),
\end{align}
so that the interaction vertex ${V}_{\bd{k}_1, \bd{k}_2; \bd{k}_3, \bd{k}_4}$
in Eq.~\eqref{eq:averageEffActionAmp} vanishes as the fourth power of the external momenta.
Obviously,  this vertex has a scaling dimension of $-2$ and is thus irrelevant for the BKT transition.
We thus conclude that 
weak amplitude fluctuations will not affect the BKT transition in a qualitative way, 
although they will slightly change non-universal quantities  like the critical temperature $T_c$. 
For a quantitative estimate, let us make the ansatz that for arbitrary cutoff scale $\Lambda$ 
amplitude fluctuations  generate a four-point vertex of the form
given by the last term in Eq.~\eqref{eq:averageEffActionAmp},
\begin{align}
 &
- \frac{g_\Lambda}{4! N} \sum_{\bd{k}_1 \bd{k}_2 \bd{k}_3 \bd{k}_4} 
 \delta_{\bd{k}_1 + \bd{k}_2 + \bd{k}_3 + \bd{k}_4, 0}
V_{\bd{k}_1, \bd{k}_2, \bd{k}_3, \bd{k}_4} 
 \nonumber
\\
 & \hspace{25mm} \times
\mb_{\bd{k}_1} \mb_{\bd{k}_2} \mb_{\bd{k}_3} \mb_{\bd{k}_4},
\end{align}
where we identify $g_{\Lambda_0} = g$. While in an exact treatment the momentum dependence of
$V_{\bd{k}_1, \bd{k}_2, \bd{k}_3, \bd{k}_4}$ will probably
change its functional form during the flow, the above ansatz
allows us to explicitly evaluate the flow of $g_\Lambda$. Introducing the rescaled coupling
\begin{align}
\tilde{g}_l = \frac{c_\Lambda \Lambda^2}{2\pi} g_\Lambda,
\label{eq:defgTilde}
\end{align}
we show in Appendix~\ref{subsec:appFlowgDeriv} that to leading order
in the fugacity the flow of $\tilde{g}_l$ is
\begin{align}
\partial_l \tilde{g}_l = -2 \tilde{g}_l - 4 \tau_l \tilde{y}_l^2.
\label{eq:gFlowEquation}
\end{align}
Replacing the scale-dependent $\tau_l$ by its initial value $\tau^{\prime}$ 
which is justified for $\tilde{y}_l \ll 1$, we can solve this flow equation for $\tilde{g}_l$,
\begin{align}
\tilde{g}_l = \tilde{g}_0 e^{-2l} - \frac{2 \tau^{\prime 2} \tilde{y}_l^2}{3 \tau^{\prime} 
- \pi} \left[ 1 - e^{-2 (3 - \pi / \tau^{\prime} ) l} \right].
\end{align}
Using the fact that for $\tau^{\prime}$  close to $\tau_\ast = \pi / 2$ 
the first term vanishes more rapidly than $\tilde{y}_l^2$, 
we conclude that the sign of $\tilde{g}_l$  changes at some intermediate $l$
and that for sufficiently large $l$ the flow of $\tilde{g}_l$ is determined by
\begin{align}
\tilde{g}_l = - \frac{2 \tau_{\ast}^2 \tilde{y}_l^2}{3 \tau_{\ast} - \pi} =  - \pi \tilde{y}_l^2.
\label{eq:gLargeL}
\end{align}
A graph of the typical RG flow of $\tilde{g}_l$ is shown in Fig.~\ref{fig:gFlow}.
The important point is that for large $l$ the flow of the irrelevant coupling $\tilde{g}_l$ is independent of its initial value $\tilde{g}_0 \propto g$ and approaches zero  at the BKT transition where the renormalized fugacity $\tilde{y}_{\ast}$ vanishes.
Note that usually  irrelevant couplings modify the RG flow of the relevant and marginal couplings only weakly so that their effect can be implicitly taken into account by redefining the numerical initial values of the relevant and marginal couplings \cite{Kopietz10}. In our case, the coupling $\tilde{g}_l$  does not change the flow equation for $\tilde{y}_l$, but it does modify the prefactor in the flow equation for $\tau_l$ (see Appendix~\ref{subsec:appFlowgEffect}),
\begin{align}
\partial_l \tau_l = \frac{\tilde{y}_l^2}{8\pi} \left( 1 + \frac{16}{3} \right).
\end{align}
However, this correction can be simply taken into account by a redefinition of $\tilde{y}_l$, as discussed in the text after Eqs.~\eqref{eq:flowtaures} and \eqref{eq:flowy4}.
We thus conclude that weak amplitude fluctuations do not modify the universal properties of the BKT transition.

\section{Effect of out-of-plane fluctuations on the BKT transition}
\label{sec:appHeisenberg}

The technique developed in the previous section can also be used to study the
effect of weak out-of-plane fluctuations of the spins on the BKT transition. 
Experimentally, it is hard to fully eliminate such fluctuations and it is therefore important so show
that they do not destroy the BKT transition.
In this section we demonstrate that weak out-of-plane fluctuations are irrelevant for the BKT transition, since they lead to the same effective field theory for the vortices as weak amplitude fluctuations.

To study the effect of out-of-plane motion of the spins,
we generalize the XY-Hamiltonian  given in Eq.~\eqref{eq:hamiltonianXY} as follows,
\begin{align}
&H_{\rm XXZ} = - J \sum_{ i , \mu  } \bd{s}_i \cdot \bd{s}_{i + \mu } + \frac{U}{2} \sum_i \left( \vartheta_i - \frac{\pi}{2} \right)^2,
\label{eq:appHamiltonianHeisenberg}
\end{align}
where the spins $\bd{s}_i$ of unit length are now allowed to rotate in 
three dimensional space, which we parametrize using spherical 
coordinates as $\bd{s}_i = \left( \sin \vartheta_i \cos \varphi_i, \sin \vartheta_i \sin \varphi_i, \cos \vartheta_i \right)$. 
The first term in Eq.~\eqref{eq:appHamiltonianHeisenberg} describes the 
classical Heisenberg model, while the second term introduces an easy-plane 
anisotropy parametrized by the coupling $U$, which for large $U$ suppresses the rotation of the spins out of the XY-plane. 
In the limit $U \to \infty$ we would recover the XY-model. 
The partition function of the above XXZ-model is
\begin{align}
&Z_{\rm XXZ} = \prod_i \left( \int d\Omega_i \right) 
\exp \bigg[ - \frac{U}{2 T} \sum_i \left( \vartheta_i - \frac{\pi}{2} \right)^2
\nonumber
\\
&+ \frac{J}{T} \sum_{i, \mu} \left[ \cos \vartheta_{i+\mu} \cos \vartheta_i + \sin \vartheta_{i+\mu} \sin \vartheta_i \cos ( \Delta_\mu \varphi_i ) \right] \bigg],
\end{align}
where $\int d\Omega_i = \int_0^{\pi} d \vartheta_i \sin \vartheta_i \int_0^{2 \pi} d \varphi_i $ denotes the integral over the unit sphere.  Similar to our treatment of amplitude fluctuations in Eq.~(\ref{eq:Villain2}),
we expand the exponential involving the cosine of the lattice gradient using 
the Villain approximation,
\begin{align}
&\indent \exp \left[ \frac{J}{T} \sin \vartheta_{i+\mu} \sin \vartheta_i \cos ( \Delta_\mu \varphi_i ) \right]
\nonumber
\\
&\approx R_{V,i\mu} \sum_{n_{i\mu} = -\infty}^\infty \exp \left[ - 
\frac{  \left( \Delta_\mu \varphi_i - 2\pi n_{i\mu} \right)^2    }{2 \tau_{V,i\mu}} \right],
\end{align}
where
\begin{subequations}
\begin{align}
R_{V,i\mu} = R_V (\sin \vartheta_{i+\mu} \sin \vartheta_i / \tau ),
\\
\tau_{V,i\mu} = \tau_V ( \sin \vartheta_{i+\mu} \sin \vartheta_i / \tau ).
 \label{eq:tauVimu}
\end{align}
\end{subequations}
Here $\tau = J / T$ and our notation is analogous to 
the notation introduced in Sec.~\ref{sec:amplitudeFlucts}. 
Performing first the integrals over  the angles $\varphi_i$ we obtain
\begin{align}
&Z_{\rm XXZ} = \prod_i \left( \int_0^\pi d\vartheta_i \sin \vartheta_i \right) \exp \bigg[ 
- \frac{U}{2 T} \sum_i \left( \vartheta_i - \frac{\pi}{2} \right)^2
\nonumber
\\
&+ \frac{J}{T} \sum_{i, \mu} \cos \vartheta_{i+\mu} \cos \vartheta_i + \sum_{i,\mu} \ln R_{V,i\mu} \bigg] Z_{\rm eff} [\vartheta],
\end{align}
where
\begin{align}
Z_{\rm eff} [\vartheta] &= \prod_i \left( \int_0^{2\pi} d\varphi_i \right) \prod_{i,\mu} \left( \sum_{n_{i\mu} = -\infty}^\infty \right)
\nonumber
\\
& \hspace{7mm} \times \exp \left[ - \frac{\left( \Delta_\mu \varphi_i - 2\pi n_{i\mu} \right)^2}{2 \tau_{V,i\mu}} \right].
\end{align}
Note that $Z_{\rm eff} [\vartheta]$ 
has the same structure as Eq.~\eqref{eq:zThetaVillainAmp}, so that we can write 
it in a current-like form similar to Eq.~\eqref{eq:ampFlucCurrentForm}. 
Defining
\begin{align}
\vartheta_i = \frac{\pi}{2} + \tTheta_i
\end{align}
we can then expand the effective action of the fields $\vartheta_i$
up to second order in $\tTheta_i$, which is justified for $U/T \gg 1$. 
The expansion of  $\tau_{V,i\mu}$ defined in Eq.~(\ref{eq:tauVimu}) is
\begin{align}
\tau_{V,i\mu} = \tau_V^{(0)} - \tau_V^{(1)} \frac{\tTheta_{i+\mu}^2 + \tTheta_i^2}{2} + \mathcal{O} \left( \tTheta_i^4 \right),
\end{align}
where the coefficients are
\begin{align}
\tau_V^{(0)} = \tau_V \left( \frac{1}{\tau} \right), \quad \tau_V^{(1)} = \frac{1}{\tau} 
\tau_V' \left( \frac{1}{\tau} \right),
\end{align}
following the notation of Eq.~\eqref{eq:tauExpansionAmp}. In the dual vortex formulation we then get
\begin{align}
Z_{\rm eff} [\vartheta] &= \exp \left[ \frac{1}{2} \sum_{i \mu} \ln \tau_{V,i\mu} \right] \prod_i \left( \sum_{m_i = - \infty}^\infty \right)
\nonumber
\\
&\times \exp \left[ - \frac{\tau_V^{(0)}}{2} \sum_i M_i
+ \frac{\tau_V^{(1)}}{2} \sum_i \tTheta_i^2 M_i \right],
\end{align}
where $M_i$ is defined in Eq.~\eqref{eq:defM}. 
Expanding $\ln R_{V,i\mu}$ and $\ln \tau_{V,i\mu}$ as well as $\sin \vartheta_i$ from the integral measure to second order in the $\tTheta_i$ only yields terms of the form $\tTheta_i^2$ with coefficients of order unity that lead to a finite renormalization
of the coupling $U$, which we absorb into the dimensionless coupling
\begin{align}
u = \frac{U}{T} + \mathcal{O} \left( 1 \right).
\end{align}
The full partition function can then be written as
\begin{align}
Z_{\rm XXZ} &= \prod_i \left( \int_{-\infty}^\infty d\tTheta_i \sum_{m_i = - \infty}^\infty \right)
\nonumber
\\
&\times
 \exp \bigg[ - \frac{\tau_V^{(0)}}{2} \sum_i M_i
 - \frac{1}{2} \sum_{i j} A_{ij} \tTheta_i \tTheta_j \bigg],
\label{eq:xxzEffZ}
\end{align}
where we have extended the range of the  $\tTheta$-integrations to the entire
real line and
\begin{align}
A_{ij} = \delta_{ij} \left( u - \tau_V^{(1)} M_i \right) - \frac{1}{\tau} \sum_\mu \left( \delta_{j,i+\mu} + \delta_{j,i-\mu} \right).
 \label{eq:Aijdef}
\end{align}
Note that $\tau_V^{(1)}$ is always negative, so that the partition function \eqref{eq:xxzEffZ} is well defined. After performing the  Gaussian integrations over the angles $\tilde{\vartheta}_i$ 
we obtain 
\begin{align}
Z_{\rm XXZ} = \prod_i \left( \sum_{m_i = - \infty}^\infty \right) \exp \bigg[ &- \frac{\tau_V^{(0)}}{2} \sum_i M_i
\nonumber
\\
&- \frac{1}{2} \ln \det \mathbf{A} \bigg],
\label{eq:lnDetG}
\end{align}
where the matrix elements of the matrix $\mathbf{A}$ 
are given in Eq.~(\ref{eq:Aijdef}).
To first order in $1/{u}$ we may neglect the off-diagonal part of $\mathbf{A}$, so that
\begin{align}
 \frac{1}{2} \ln \det \mathbf{A} =  \frac{1}{2} \sum_i \ln \left[ u - \tau_V^{(1)} M_i \right].
 \label{eq:logDetADiag}
\end{align}
Finally, expanding the logarithm up to first order in $1/{u}$ we obtain
\begin{align}
&Z_{\rm XXZ} = \prod_i \left( \sum_{m_i = - \infty}^\infty \right) \exp \left[ - \frac{
 \tau^{\prime}}{2} \sum_i M_i \right],
\end{align}
which is just the partition function of the XY-model in the dual vortex formulation with renormalized coupling
\begin{align}
\tau^{\prime} = \tau_V^{(0)} - \frac{\tau_V^{(1)}}{u}.
\end{align}
Relaxing the confinement of the spins to the XY-plane can thus to leading order be fully absorbed in a redefinition of the dimensionless temperature in the dual vortex action of the XY-model. Considering also terms of order $1/u^2$, we find that the off-diagonal part of $\mathbf{A}$ only contributes an unimportant constant term. Expanding the logarithm \eqref{eq:logDetADiag} as well as the resulting exponential to second order in $1/u$ then yields
\begin{align}
Z_{\rm XXZ} &= \prod_i \left( \sum_{m_i = - \infty}^\infty \right) \exp \left[ - \frac{\tau^{\prime}}{2} \sum_i M_i \right]
\nonumber
\\
&\hspace{20mm} \times \left( 1 + \frac{g}{4!} \sum_i M_i^2 \right),
\label{eq:xxzZ}
\end{align}
where
\begin{align}
g = \frac{6 \bigl( \tau_V^{(1)} \bigr)^2}{u^2}.
\end{align}
Since the expression \eqref{eq:xxzZ} for $Z_\text{XXZ}$ is formally identical to the partition function \eqref{eq:ampZ}, weak out-of-plane fluctuations lead to the same effective field theory as weak amplitude fluctuations. Obviously, the results in Sec.~\ref{subsec:ampFlowEqs} therefore also apply to the strongly anisotropic XXZ-model, so that we conclude that the BKT transition is not destroyed by a small out-of-plane
motion of the spins.

\section{Summary and conclusions}

In this work we have shown how to recover the
Kosterlitz-Thouless RG flow equations
for the two-dimensional XY-model entirely
within the framework of the FRG. The crucial technical step is to apply the
lattice version of the 
FRG formalism developed by Machado and Dupuis \cite{Machado10}
to a dual formulation of the XY-model, where integer degrees of freedom 
represent the vortices associated with  the sites of the dual lattice.
With this method, the Kosterlitz-Thouless RG flow equations
can be obtained by means of an almost automatic application of the
established FRG machinery in momentum space.
The only technical complication encountered in this approach is that
one has to solve an infinite hierarchy of  RG flow equations for the 
relevant and marginal coupling constants of the model.
Fortunately, to leading order in the fugacity,
an explicit solution of this infinite hierarchy can be found, as shown in 
Sec.~\ref {sec:BKTinXYfromFRG}.
In contrast to previous attempts to recover the BKT transition
within the framework of the FRG \cite{Graeter95,Gersdorff01,Jakubczyk14,Jakubczyk17},
with our dual lattice FRG
we exactly recover the 
RG flow equations obtained by Kosterlitz and Thouless \cite{Kosterlitz73}, which are believed
to correctly describe
the critical properties of the BKT transition.
Our dual lattice FRG in momentum space is therefore an alternative
to the rather unconventional real space RG used by Kosterlitz and Thouless
in their original work \cite{Kosterlitz73}.

An important advantage of our FRG approach is that 
it can be easily generalized to 
take into account various deviations from the ideal XY-model,
such as amplitude fluctuations (see Sec.~\ref{sec:amplitudeFlucts}) 
or small out-of-plane movements of the spins (see Sec.~\ref{sec:appHeisenberg}).
In particular, we have shown that
amplitude as well as out-of-plane fluctuations 
give rise to an irrelevant effective interaction between  vortices, 
which does not qualitatively modify the BKT transition. 
This is in contrast to previous FRG approaches \cite{Graeter95,Gersdorff01,Jakubczyk14,Jakubczyk17}, which did  not explicitly take 
the vortex degrees of freedom into account and did not 
obtain the line of true fixed points associated with the BKT phase 
without fine-tuning the regulator.

\begin{acknowledgments}
We thank Henk Hilhorst for pointing out that the dual representation of the partition function of the Villain model given in Eq.~\eqref{eq:dual1} is actually divergent. This work was supported by the SFB Transregio 49
of the Deutsche Forschungsgemeinschaft.
\end{acknowledgments}

\begin{appendix}

\appendix

\section{TECHNICAL DETAILS ON FLOW EQUATIONS FOR THE XY-MODEL}
\label{sec:appTechDetails}

\subsection{Flow equations for the marginal couplings}
\label{sec:appFlowMarginalAllOrders}

In Sec.~\ref{sec:xyFlowMarginal} we have shown that for $n = 2,3,4$, the flow of the marginal couplings $c^{(2n)}_\Lambda$ is to leading order in $y_\Lambda$ given by
\begin{equation}
   \partial_{l} c_{l}^{(2n)} =  \frac{ c_{l}^{(2n+2)}}{ 4 \pi
 \tau_{l}} + (-1)^n ( 4 \pi)^{2n-3}  \tilde{y}_l^2,
 \label{eq:appFlowMarginal}
 \end{equation}
see Eq.~\eqref{eq:cnflowl}. In the following we will show that this equation holds for all integers $n \geq 2$. Consider first the right-hand side of Eq.~\eqref{eq:appFlowMarginal}, which according to Eq.~\eqref{eq:resultMarginalGeneral} is proportional to $\tilde{y}_l^2$. Since all vertices $\Gamma^{(2n)}_\Lambda$ for $n \geq 2$ are at least of order $\tilde{y}_l$, it follows that for the flow equation of $c_l^{(2n)}$ we only have to consider terms with up to two vertices, so that we can write
\begin{align}
\partial_\Lambda \Gamma^{(2n)}_\Lambda (\bd{k}) 
= \dot{\Gamma}^{(2n,1)}_\Lambda (\bd{k}) + \dot{\Gamma}^{(2n,2)}_\Lambda (\bd{k}),
\label{eq:appVertexGammas}
\end{align}
where $\dot{\Gamma}^{(2n,m)}_\Lambda (\bd{k})$ 
only contains terms with exactly $m$ vertices. For $\dot{\Gamma}^{(2n,1)}_\Lambda (\bd{k})$ we find \cite{Kopietz10}
\begin{align}
\dot{\Gamma}^{(2n,1)}_\Lambda (\bd{k}) &= \frac{1}{2N} \sum_{\bd{q}} \dot{G}_\Lambda (\bd{q}) \Gamma^{(2n+2)}_\Lambda (\bd{q},-\bd{q},\bd{k},-\bd{k},0, \ldots, 0)
\nonumber
\\
&= \frac{A_\Lambda}{2} \Gamma^{(2n+2)}_\Lambda (\bd{k}) + \mathcal{O} (\tilde{y}_l),
\end{align}
where $A_{\Lambda}$ is defined in Eq.~(\ref{eq:Adef}) and
we have used the fact that to leading order in $\tilde{y}_l$
we may neglect the loop momentum $\bd{q}$ in the vertex on the right-hand side. 

Next, consider the second contribution
 $\dot{\Gamma}^{(2n,2)}_\Lambda (\bd{k})$ 
in Eq.~(\ref{eq:appVertexGammas}); all contributions to this term 
are of the form \cite{Kopietz10}
\begin{align}
& \frac{1}{N} \sum_{\bd{p} \bd{q}} \dot{G}_\Lambda (\bd{q}) G_\Lambda (\bd{p}) \Gamma^{(n_1 + 2)}_\Lambda (\bd{q},-\bd{p},\bd{k}_1, \ldots, \bd{k}_{n_1})
\nonumber
\\
&\times \Gamma^{(n_2 + 2)}_\Lambda (\bd{p},-\bd{q},\bd{k}_{n_1+1}, \ldots, \bd{k}_{2n}),
\label{eq:genFormM2}
\end{align}
where $n_1$ and $n_2$ are positive even integers which fulfill $n_1 + n_2 = 2n$ and the set of momenta $\{\bd{k}_1, \ldots, \bd{k}_{2n}\}$ is a permutation of $\{ \bd{k}, -\bd{k}, 0, \ldots, 0\}$. Since all permutations with $\bd{k}$ and $-\bd{k}$ belonging to the same vertex contribute at least to third order in $\tilde{y}_l$ to the flow of $c^{(2n)}_\Lambda$, we only have to consider the cases where $\bd{k}$ and $-\bd{k}$ belong to different vertices. Approximating the vertices by their momentum-independent parts and explicitly considering the combinatorial factors then yields
\begin{align}
\dot{\Gamma}^{(2n,2)}_\Lambda (\bd{k}) &= - \frac{1}{2} \sum_{n_1 = 1}^\infty \sum_{n_2 = 1}^\infty \delta_{n_1+n_2,2n} \frac{2 n_1 n_2 (2n-2)!}{n_1! n_2!}
\nonumber
\\
&\times u^{(n_1 + 2)}_\Lambda u^{(n_2 + 2)}_\Lambda B_\Lambda (k),
\end{align}
where we have used $B_\Lambda (k)$ from Eq.~\eqref{eq:definitionBLambdaK}. The combinatorial factors in the numerator correspond to the symmetry $\bd{k} \leftrightarrow -\bd{k}$, to the $n_1$ ($n_2$) possibilities to put $\bd{k}$ or $-\bd{k}$ in the first (second) vertex, and to the $(2n-2)!$ possibilities of distributing the remaining (vanishing) momenta, while the denominator is due to the $n_1!$ ($n_2!$) possible permutations of the indices in the first (second) vertex. Since in our system vertices with an odd number of external legs vanish, we can rewrite this using binomial coefficients as
\begin{align}
\dot{\Gamma}^{(2n,2)}_\Lambda (\bd{k}) = - \sum_{i = 1}^{n-1}
\begin{pmatrix}
2n - 2 \\ 2i - 1
\end{pmatrix}
u^{(2i + 2)}_\Lambda u^{(2n - 2i + 2)}_\Lambda B_\Lambda (k).
\end{align}
With Eqs.~(\ref{eq:undef}) and (\ref{eq:tildey}) we find
\begin{align}
u^{(2i + 2)}_\Lambda u^{(2n - 2i + 2)}_\Lambda = (-1)^n c_\Lambda^2 \Lambda^4 (2\pi)^{2n-2} \tilde{y}_l^2,
\label{eq:uuWithY}
\end{align}
which is manifestly independent of the summation index $i$. This allows us to perform the sum so that
\begin{align}
\dot{\Gamma}^{(2n,2)}_\Lambda (\bd{k}) = - 2^{2n - 3}
u^{(4)}_\Lambda u^{(2n)}_\Lambda B_\Lambda (k).
\end{align}
Acting with $\lim_{k \to 0} \partial_k^2$ on our results for 
$\dot{\Gamma}^{(2n,1)}_\Lambda (\bd{k})$ and $\dot{\Gamma}^{(2n,2)}_\Lambda (\bd{k})$, inserting our previous leading order results for $A_\Lambda$ and $B_\Lambda''$, and using the logarithmic scale derivative $\partial_l = - \Lambda \partial_\Lambda$ then results in Eq.~\eqref{eq:appFlowMarginal}.

\subsection{Flow equations for the relevant couplings}
\label{sec:appFlowRelevantQuadraticY}

In the main text we have derived the flow equations for the relevant couplings $u_\Lambda^{(2n)}$ only to leading order in $y_\Lambda$ [cf. Eq.~\eqref{eq:flowuna}],
\begin{equation}
  \partial_{l} u_{\Lambda}^{(2n)}  =  
 \frac{u_{\Lambda}^{(2n+2)}}{ 4 \pi \tau_\Lambda} +
 {\cal{O}} ( y_{\Lambda}^2 ).
 \end{equation}
In the following we will extend this result by also taking second-order terms in $y_\Lambda$ into account. Using the notation introduced in Eq.~\eqref{eq:appVertexGammas}, we need to calculate
\begin{align}
\partial_\Lambda \Gamma^{(2n)}_\Lambda (0) = 
\dot{\Gamma}^{(2n,1)}_\Lambda (0) + \dot{\Gamma}^{(2n,2)}_\Lambda (0)
 \label{eq:dotGamma2n}
\end{align}
up to order $y_\Lambda^2$. For the first term we find \cite{Kopietz10}
\begin{align}
\dot{\Gamma}^{(2n,1)}_\Lambda (0) = \frac{1}{2N} \sum_{\bd{q}} \dot{G}_\Lambda (\bd{q}) \Gamma^{(2n+2)}_\Lambda (\bd{q}),
\end{align}
where we now have to take the loop momentum $\bd{q}$ in the vertex into account. Inserting our long-wavelength expansion for the vertices in Eq.~\eqref{eq:vertexexp} yields
\begin{align}
\dot{\Gamma}^{(2n,1)}_\Lambda (0) = \frac{u^{(2n+2)}_\Lambda}{2} A_\Lambda + \frac{a^2 c_\Lambda^{(2n+2)}}{4} A^{(2)}_\Lambda,
\end{align}
where we have defined
\begin{align}
A^{(2)}_\Lambda = \frac{1}{N} \sum_{\bd{q}} \dot{G}_\Lambda (\bd{q}) q^2.
\end{align}
From Eq.~\eqref{eq:Adef} we see that $A_\Lambda$ is to first order in $y_\Lambda$ given by
\begin{align}
A_\Lambda = - \frac{1}{2\pi \tau_\Lambda \Lambda} + \frac{u_\Lambda^{(2)}}{\pi \tau_\Lambda c_\Lambda \Lambda^3} + \mathcal{O} (y_\Lambda^2),
\end{align}
while for $A^{(2)}_\Lambda$ we only need the leading order result
\begin{align}
A^{(2)}_\Lambda = - \frac{\Lambda}{4\pi \tau_\Lambda} + \mathcal{O} (y_\Lambda),
\end{align}
since $c^{(2n+2)}_\Lambda$ is already of second order in $y_\Lambda$. Thus
\begin{align}
- \Lambda \dot{\Gamma}^{(2n,1)}_\Lambda (0) = \frac{u^{(2n+2)}_\Lambda}{4\pi \tau_\Lambda} - \frac{u_\Lambda^{(2)} u^{(2n+2)}_\Lambda}{2 \pi \tau_\Lambda c_\Lambda \Lambda^2} + \frac{a^2 \Lambda^2 c_\Lambda^{(2n+2)}}{16\pi \tau_\Lambda}.
\end{align}

Next, consider the second contribution  $\dot{\Gamma}^{(2n,2)}_\Lambda (0)$ 
in Eq.~(\ref{eq:dotGamma2n}), which can be written as \cite{Kopietz10}
\begin{align}
\dot{\Gamma}^{(2n,2)}_\Lambda (0) &= - \frac{1}{2N} \sum_{n_1 = 1}^\infty \sum_{n_2 = 1}^\infty \delta_{2n,n_1 + n_2} \frac{(2n)!}{n_1! n_2!}
\nonumber
\\
&\times \sum_{\bd{q}} \dot{G}_\Lambda (\bd{q}) G_\Lambda (\bd{q}) \Gamma^{(n_1+2)}_\Lambda (\bd{q}) \Gamma^{(n_2+2)}_\Lambda (\bd{q}),
\end{align}
where the combinatorial factors correspond to the $(2n)!$ possible permutations of the $2n$ external vanishing momenta and to the $n_1!$ ($n_2!$) possibilities to distribute $n_1$ ($n_2$) momenta on the first (second) vertex. Since this expression is already of order $y_\Lambda^2$, we can neglect the $\bd{q}$ dependence of the vertices. Using the definition of $B_\Lambda (\bd{k})$ in Eq.~\eqref{eq:definitionBLambdaK} as well as the fact that vertices with an odd number of external legs vanish identically, we find
\begin{align}
\dot{\Gamma}^{(2n,2)}_\Lambda (0) &= - \frac{B_\Lambda (0)}{2} \sum_{i = 1}^{n-1}
\begin{pmatrix}
2n \\ 2i
\end{pmatrix}
u^{(2i+2)}_\Lambda u^{(2n-2i+2)}_\Lambda.
\end{align}
Inserting $B_\Lambda (0)$ from Eq.~\eqref{eq:BLambdaK0} we can also write this as
\begin{align}
- \Lambda \dot{\Gamma}^{(2n,2)}_\Lambda (0) &= - \sum_{i = 1}^{n-1}
\begin{pmatrix}
2n \\ 2i
\end{pmatrix}
\frac{u^{(2i+2)}_\Lambda u^{(2n-2i+2)}_\Lambda}{4\pi \tau_\Lambda c_\Lambda \Lambda^2}.
\end{align}
Combining our results for $\dot{\Gamma}^{(2n,1)}_\Lambda (0)$ 
and $\dot{\Gamma}^{(2n,2)}_\Lambda (0)$ then leads to
\begin{align}
\partial_l u_\Lambda^{(2n)} &= \frac{u^{(2n+2)}_\Lambda}{4\pi \tau_\Lambda} + \frac{a^2 \Lambda^2 c_\Lambda^{(2n+2)}}{16\pi \tau_\Lambda}
\nonumber
\\
&- \sum_{i = 0}^n
\begin{pmatrix}
2n \\ 2i
\end{pmatrix}
\frac{u^{(2i+2)}_\Lambda u^{(2n-2i+2)}_\Lambda}{4\pi \tau_\Lambda c_\Lambda \Lambda^2} + \mathcal{O} (y_\Lambda^3).
\end{align}
Since Eq.~\eqref{eq:uuWithY} is correct to second order in $\tilde{y}_l$, we can perform the sum over $i$ explicitly to obtain
\begin{align}
\partial_l u_\Lambda^{(2n)} &= \frac{u^{(2n+2)}_\Lambda}{4\pi \tau_\Lambda} + \frac{a^2 \Lambda^2 c_\Lambda^{(2n+2)}}{16\pi \tau_\Lambda}
\nonumber
\\
&- \frac{(-1)^n a^2 \Lambda^2 (4\pi)^{2n-2} \tilde{y}_l^2}{2\pi} + \mathcal{O} (y_\Lambda^3),
\label{eq:flowUSecondOrder}
\end{align}
which is valid for all $n \geq 1$.

\subsection{$\tilde{y}_l^2$ correction to the flow of the fugacity}
\label{sec:appNextToLeadingOrderFugacity}

We can now use our result \eqref{eq:flowUSecondOrder} to calculate the $\tilde{y}_l^2$ correction at $\tau = \tau_\ast$ to the flow of the fugacity. For this purpose, we parametrize the relevant couplings $u_\Lambda^{(2n)}$ as
\begin{align}
u_l^{(2n)} = (-1)^{n+1} (2\pi)^{2n-3} \tau_l a^2 \Lambda^2 \tilde{y}_l \left( 1 + 4^{n-2} \upsilon_l^{(2n)} \tilde{y}_l \right),
\end{align}
where $\upsilon_l^{(2)} = 0$ by construction. We then
insert our result \eqref{eq:resultMarginalGeneral} for $c_\Lambda^{(2n)}$ into Eq.~\eqref{eq:flowUSecondOrder} and approximate $\tau_l = \tau_\ast$ in all higher order terms. For $n=1$ this yields
\begin{align}
\frac{\partial_l \tilde{y}_l}{\tilde{y}_l} = 2 - \frac{\pi}{\tau_l} - \frac{\pi}{\tau_l} \left( \upsilon_l^{(4)} - \frac{9}{8\pi} \right) \tilde{y}_l,
\label{eq:flowYSecondOrder}
\end{align}
while for $n \geq 2$ we find
\begin{align}
\partial_l \upsilon_l^{(2n)} &= -2 \left( 4 \upsilon_l^{(2n+2)} - \upsilon_l^{(2n)} - 4^{2-n} \upsilon_l^{(4)} \right)
\nonumber
\\
&+ \frac{9}{\pi} \left( 1 - 4^{1-n} \right).
\label{eq:flowUpsilon2nGeneral}
\end{align}
This infinite hierarchy of flow equations can be solved with the ansatz
\begin{align}
\frac{\upsilon_l^{(2n+2)}}{\upsilon_l^{(2n)}} = \frac{1 - 4^{-n}}{1 - 4^{1-n}},
\label{eq:ansatzUpsilon}
\end{align}
which implies
\begin{align}
\upsilon_l^{(4)} = \frac{3}{4 (1 - 4^{1-n})} \upsilon_l^{(2n)}.
\end{align}
Inserting our ansatz into Eq.~\eqref{eq:flowUpsilon2nGeneral} yields
\begin{align}
\partial_l \upsilon_l^{(2n)} &= -6 \upsilon_l^{(2n)} + \frac{9}{\pi} \left( 1 - 4^{1-n} \right),
\end{align}
which is easily solved as
\begin{align}
\upsilon_l^{(2n)} = \frac{3 (1 - 4^{1-n})}{2\pi} (1 - e^{-6l}),
\end{align}
justifying our ansatz \eqref{eq:ansatzUpsilon}. For $n=2$ and large $l$ we find
\begin{align}
\upsilon_l^{(4)} = \frac{9}{8\pi},
\end{align}
which exactly cancels the linear in $\tilde{y}_l$ contribution 
in Eq.~\eqref{eq:flowYSecondOrder}.

\section{ADDITIONAL FLOW EQUATION DUE TO AMPLITUDE FLUCTUATIONS}
\label{sec:appFlowg}

\subsection{Flow equation for $g_\Lambda$}
\label{subsec:appFlowgDeriv}

In the following we will derive the flow equation for the quartic coupling $g_\Lambda$, which results from integrating out the amplitude fluctuations. To this end we approximate all vertices by their small-momentum expansion from Eq.~\eqref{eq:vertexexp},
\begin{equation}
\Gamma^{(2n)}_{\Lambda} ( \bd{k} , - \bd{k} , 0 , \ldots , 0 ) =
u_{\Lambda}^{(2n)} +  \frac{ k^2 a^2 }{2} c^{(2n)}_{\Lambda}    + {\cal{O}} ( k^4 ),
\end{equation}
except for the four-point vertex where we additionally include
\begin{align}
 & - \frac{g_\Lambda}{4! N} \sum_{\bd{k}_1 \bd{k}_2 \bd{k}_3 \bd{k}_4} 
 \delta_{\bd{k}_1 + \bd{k}_2 + \bd{k}_3 + \bd{k}_4, 0}
V_{\bd{k}_1, \bd{k}_2, \bd{k}_3, \bd{k}_4} 
 \nonumber
 \\
 & \hspace{20mm} \times
\mb_{\bd{k}_1} \mb_{\bd{k}_2} \mb_{\bd{k}_3} \mb_{\bd{k}_4}
\end{align}
in our ansatz for the average effective action, since the initial value $g_{\Lambda_0} = g$ of this fourth-order momentum term is finite (cf. Sec.~\ref{sec:amplitudeFlucts}). Using the long-wavelength limit
\begin{align}
V_{\bd{k},\bd{k},-\bd{k},-\bd{k}} \approx a^4 k^4
\end{align}
as well as the exact flow equation \eqref{eq:flow4}  for the four-point vertex, we see that
\begin{align}
\partial_\Lambda g_\Lambda &= - \frac{1}{4! a^4} \lim_{k \to 0} \partial_k^4 \partial_\Lambda \Gamma^{(4)}_{\Lambda,\bd{k},\bd{k},-\bd{k},-\bd{k}} = \frac{1}{4! a^4 N}
\nonumber
\\
&\times \sum_{\bd{p}} \dot{G}_\Lambda (\bd{p}) \lim_{k \to 0} \partial_k^4 \bigg[ 2 G_\Lambda (\bd{p}) \left( \Gamma^{(4)}_{\Lambda,\bd{p},-\bd{p},\bd{k},-\bd{k}} \right)^2
\nonumber
\\
&+ G_\Lambda (\bd{p} + 2\bd{k}) \left( \Gamma^{(4)}_{\Lambda,\bd{p},-\bd{p}-2\bd{k},\bd{k},\bd{k}} \right)^2 \bigg].
\end{align}
Keeping only leading order terms in $y_\Lambda$ and $g_\Lambda$, this can be written more explicitly as
\begin{align}
\partial_\Lambda g_\Lambda &= \frac{1}{4! a^4 N} \sum_{\bd{p}} \dot{G}_\Lambda (\bd{p}) \lim_{k \to 0} \bigg[ \left( u_\Lambda^{(4)} \right)^2 \partial_k^4 G_\Lambda (\bd{p} + 2\bd{k})
\nonumber
\\
&- 12 u_\Lambda^{(4)} g_\Lambda \left( \partial_k^2 G_\Lambda (\bd{p} + 2\bd{k}) \right) \partial_k^2 V_{\bd{p},-\bd{p}-2\bd{k},\bd{k},\bd{k}}
\nonumber
\\
&+ 18 g_\Lambda^2 G_\Lambda (\bd{p}) \left( \partial_k^2 V_{\bd{p},-\bd{p},\bd{k},-\bd{k}} \right)^2 \bigg],
\end{align}
where we can use the long-wavelength expansion
\begin{align}
V_{\bd{p},-\bd{p},\bd{k},\pm \bd{k}} \approx \mp \frac{a^4 p^2 k^2}{3} \left( 1 + 2 \cos^2 \varphi_p \right).
\label{eq:longWavelengthVDiff}
\end{align}
With
\begin{align}
&\frac{1}{N} \sum_{\bd{p}} \dot{G}_\Lambda (\bd{p}) G_\Lambda (\bd{p}+2\bd{k}) = - \frac{1}{2\pi \tau_\Lambda^2 a^2 \Lambda^3}
\nonumber
\\
&\times \left[ 1 - 2 \frac{k^2}{\Lambda^2} + \frac{80}{9\pi} \frac{k^3}{\Lambda^3} - 4 \frac{k^4}{\Lambda^4} + \mathcal{O} (k^5) \right] + \mathcal{O} (\tilde{y}_l)
\end{align}
and
\begin{align}
&\frac{1}{N} \sum_{\bd{p}} p^2 \left( 1 + 2 \cos^2 \varphi_p \right) \dot{G}_\Lambda (\bd{p}) \lim_{k \to 0} \partial_k^2 G_\Lambda (\bd{p} + 2\bd{k})
\nonumber
\\
&= \frac{5}{\pi \tau_\Lambda^2 a^2 \Lambda^3} + \mathcal{O} (\tilde{y}_l)
\end{align}
we find in total
\begin{align}
\partial_l g_\Lambda &= - \frac{8\pi \tilde{y}_l^2}{a^2 \Lambda^2} + \frac{10 \tilde{y}_l g_\Lambda}{3 \tau_l} + \frac{a^2 \Lambda^2 g_\Lambda^2}{4\pi \tau_\Lambda^2}
\nonumber
\\
&+ \mathcal{O} (\tilde{y}_l^3, \tilde{y}_l^2 g_\Lambda, \tilde{y}_l g_\Lambda^2, g_\Lambda^3).
\end{align}
Rewriting this in terms of the rescaled coupling $\tilde{g}_l$ [see Eq.~\eqref{eq:defgTilde}] we arrive at
\begin{align}
\partial_l \tilde{g}_l = \left( -2 + \frac{10 \tilde{y}_l}{3 \tau_l} + \frac{\tilde{g}_l}{2 \tau_l^3} \right) \tilde{g}_l - 4 \tau_l \tilde{y}_l^2.
\end{align}
Finally, we note that due to $\tilde{y}_l \ll 1$ and $\tilde{g}_l \ll 1$ we may neglect the second and third term in the brackets above, so that
\begin{align}
\partial_l \tilde{g}_l = -2 \tilde{g}_l - 4 \tau_l \tilde{y}_l^2.
\end{align}

\subsection{Effect of $\tilde{g}_l$ on the flow of $\tau_l$}
\label{subsec:appFlowgEffect}

Consider now the exact flow equation \eqref{eq:cexactflow} for the scale-dependent dimensionless temperature,
\begin{align}
\partial_{\Lambda} \tau_{\Lambda} &= \frac{1}{4 a^2 N} \sum_{\bd{q}} \dot{G}_{\Lambda} ( \bd{q} ) \lim_{ k \rightarrow 0} \partial_k^2 \Gamma^{(4)}_{\Lambda}  ( \bd{k} ,  - \bd{k} ,  \bd{q} , -  \bd{q}).
\end{align}
Inserting our earlier result \eqref{eq:flowtaures} for $\tau_\Lambda$ close to the BKT transition, which was derived without taking the irrelevant coupling $g_\Lambda$ into account, we find
\begin{align}
\partial_\Lambda \tau_\Lambda &= - \frac{\tilde{y}_\Lambda^2}{8\pi \Lambda} - \frac{g_\Lambda}{4 a^2 N} \sum_{\bd{q}} \dot{G}_{\Lambda} ( \bd{q} ) \lim_{ k \rightarrow 0} \partial_k^2 V_{\bd{k},-\bd{k},\bd{q},-\bd{q}}
\nonumber
\\
&= - \frac{\tilde{y}_\Lambda^2}{8\pi \Lambda} + \frac{g_\Lambda a^2 \Lambda}{12\pi \tau_\Lambda},
\label{eq:flowTauGLambda}
\end{align}
where we have used the long-wavelength expansion \eqref{eq:longWavelengthVDiff} of $V_{\bd{k},-\bd{k},\bd{q},-\bd{q}}$ as well as
\begin{align}
\frac{1}{N} \sum_{\bd{q}} \dot{G}_{\Lambda} ( \bd{q} ) q^2 \left( 1 + 2 \cos^2 \varphi_q \right) = - \frac{\Lambda}{2\pi \tau_\Lambda} + \mathcal{O} \left( \tilde{y}_\Lambda \right).
\end{align}
With the definition \eqref{eq:defgTilde} of the rescaled coupling $\tilde{g}_l$ and the logarithmic scale derivative $l = \ln (\Lambda_0 / \Lambda)$ we can also write Eq.~\eqref{eq:flowTauGLambda} as
\begin{align}
\partial_l \tau_l = \frac{\tilde{y}_l^2}{8\pi} - \frac{\tilde{g}_l}{6 \tau_l^2}.
\end{align}
Since we are interested in the flow close to the BKT transition, we may approximate $\tau_l \approx \tau_\ast = \pi / 2$ and insert our earlier result \eqref{eq:gLargeL} for $\tilde{g}_l$, so that we finally get
\begin{align}
\partial_l \tau_l = \frac{\tilde{y}_l^2}{8\pi} + \frac{2 \tilde{y}_l^2}{3 \pi} = \frac{\tilde{y}_l^2}{8\pi} \left( 1 + \frac{16}{3} \right).
\end{align}

\end{appendix}

\bibliographystyle{apsrev4-1}
\bibliography{bkt}

\end{document}